\documentclass[journal,twoside,web]{ieeecolor}

\usepackage{generic}
\usepackage{cite}
\usepackage{amsmath,amssymb,amsfonts}
\usepackage{mathrsfs}
\usepackage{lipsum}
\usepackage{algorithmic}
\usepackage{graphicx}
\usepackage{textcomp}
\usepackage[ruled,vlined]{algorithm2e}
\usepackage{tabularx}
\usepackage{array}

\newtheorem{theorem}{Theorem}
\newtheorem{lemma}{Lemma}
\newtheorem{definition}{Definition}

\newtheorem{assumption}{Assumption}
\newtheorem{proposition}{Proposition}

\newcommand{\argmin}{\mathop{\arg\min}}

\newcommand{\bigroman}[1]{\uppercase\expandafter{\romannumeral#1}}

\DeclareMathOperator{\var}{Var}

\begin{document}

\title{Convergence Theory of Generalized Distributed Subgradient Method with Random Quantization}

\author{Zhaoyue Xia, \IEEEmembership{Student Member, IEEE},
        Jun Du, \IEEEmembership{Senior Member, IEEE}, Yong Ren, \IEEEmembership{Senior Member, IEEE}
        %H. Vincent Poor, \IEEEmembership{Fellow, IEEE} % <-this % stops a space
        %Lajos~Hanzo,~\IEEEmembership{Life~Fellow,~IEEE}
\thanks{Z. Xia, J. Du, and Y. Ren are with the Department of Electronic Engineering, Tsinghua University, Beijing, 100084, China, (e-mail: xiazy19@mails.tsinghua.edu.cn; \{jundu,reny\}@mail.tsinghua.edu.cn).}} % <-this % stops a space
%\thanks{H. Vincent Poor is with the Department of Electrical Engineering, Princeton University, Princeton, NJ 08544 USA, Email: poor@princeton.edu.}}

\maketitle

\begin{abstract}
The distributed subgradient method (DSG) is a widely discussed algorithm to cope with large-scale distributed optimization problems in the arising machine learning applications. Most exisiting works on DSG focus on ideal communication between the cooperative agents such that the shared information between agents is exact and perfect. This assumption, however, could lead to potential privacy concerns and is not feasible when the wireless transmission links are not of good quality. To overcome the challenge, a common approach is to quantize the data locally before transmission, which avoids exposure of raw data and significantly reduces the size of data. Compared with perfect data, quantization poses fundamental challenges on loss of data accuracy, which further impacts the convergence of the algorithms. To settle the problem, we propose a generalized distributed subgradient method with random quantization. We provide comprehensive results on the convergence of the algorithm for (strongly/weakly) convex objective functions. We also derive upper bounds on the convergence rates in terms of the quantization bit, stepsizes and the number of network agents. Our analysis extends the existing results, where special cases of stepsizes and convex objective functions are considered, to general conclusions on non-convex cases. Finally, numerical simulations are conducted on convex and non-convex settings to support our theoretical results. 
\end{abstract}

\begin{IEEEkeywords}
Distributed consensus, distributed optimization, nonconvex optimization, multiagent systems.
\end{IEEEkeywords}

\section{Introduction}
Distributed optimization in multiagent networks has been receiving widespread research interest in the past few years with application to wireless sensor networks, robotic networks, and social networks \cite{2016WirelessCommuUAV}. There are a wide range of problems which can be formulated via distributed optimization such as cooperative detection, target tracking, and edge learning. Different from a (semi-)centralized manner, agents in fully distributed networks only share information with local neighboring agents instead of a public center. This significantly promotes efficiency of local decision making and the system performance suffers much less from the growth of the number of agents than a centralized network. Moreover, the decentralized techniques have gained popularity due to the robustness against local failures, the remarkable cut on transmission overheads of individual agents and the modest privacy protection for local data. In terms of efficiency promotion and privacy issues, it is of considerable importance to use distributed optimization to cope with large-scale distributed networks while handling substanstial amounts of generated data \cite{2016FogIoT}. In such distributed networks, the agents cooperatively accomplish a task, which is equivalent to training a global model and optimizing a global function correpsonding to the specific task. The global function is composed of local functions, where each local function is only available to the corresponding individual agent. Since there is no central connection between agents in the network, they can only share local information with neighboring nodes and cooperatively optimize the global function. 

Typical solutions to distributed optimization problems, which have been widely investigated recently, can be categorized into three classes in general, namely, the alternating direction method of multipliers (ADMM) \cite{2022LowComplexityADMM}, distributed primal-dual methods \cite{2022DistributedPrimalDual}, and distributed subgradient methods (DSG). ADMM is an algorithm that solves convex optimization problems by dividing them into smaller subproblems. Since the subproblems are easier to solve, the solutions can be obtained via iteration. Distributed primal-dual methods are proposed to deal with constrained optimization problems. Primal-dual algorithms are conducted in the augmented Lagrangian framework, which generally requires the separability of globally coupled constraints. In this article, we restrict our discussion to DSG since DSG methods are convenient to demonstrate and have stronger convergence guarantees than those for ADMM and primal-dual methods. For more information on DSG, it refers to \cite{2009DSG,2018NetworkTopology} on its properties, applications and extensions.

Throughout this article, we focus on developing a theoretical convergence result on the generalized distributed subgradient methods (GDSRQ), which is generalization to existing results reported in \cite{2021ConvergenceRates}. Different from a typical DSG method, the communication overhead limitation is considered in the scheme. That is, the shared information between agents are firstly quantized locally at a fixed bit length, and then transmitted over wireless links. In the scheme, one stepsize $\alpha_k$ is used to control the subgradient, and another stepsize $\beta_k$ indicates the mixing rate between the local parameter of an individual agent and the received parameters from its neighboring ones. We will show that under mild conditions and via appropriate choices of the two stepsizes, strong convergence results are available with respect to the optimal solution. 

\subsection{Main Contributions}
The main contributions of the article can be summarized as follows. We firstly propose the generalized distributed subgradient method as a generalization to \cite{2021ConvergenceRates}. This method uses two different stepsizes to control the process of gradient descent and parameter mixing between the local agent and neighboring ones. Then we present different covnergence results with respect to convex and non-convex objective functions. 

\textbf{Convex Cases}: We derive the sufficient conditions for the almost sure convergence of local parameters $\mathbf{x}_{i,k}$ and the corresponding weighted time-average parameters $\mathbf{z}_{i,k}$. In \cite{2021ConvergenceRates}, the convergence results of weighted time-average parameters are established on specific forms of time-average weights. To extend the results, we derive general conclusions on the convergence of weighted time-average parameters by showing that the function value at $\mathbf{z}_{i,k}$ approches the theoretically optimal value $f^*$ with respect to iteration $k$ for both strongly convex objective functions and convex objective functions. In addition, we provide the explicit form of convergence rates when the two stepsizes $\alpha_k$, $\beta_k$, and the time-average weights $\gamma_k^t$ are given by $\alpha_0(k+1)^{-\lambda_\alpha}$, $\beta_0(k+1)^{-\lambda_\beta}$, and $\frac{(t+1)^{-\lambda_\gamma}}{\sum_{t=0}^k (t+1)^{-\lambda_\gamma}}$ respectively. We find that the convergence rate of $\mathbf{z}_{i,k}$ is independent of the time-average weights $\gamma_k^t$, which means the convergence rate of $\mathbf{z}_{i,k}$ precisely reflects the covnergence rate of the local parameter $\mathbf{x}_{i,k}$. 

\textbf{Weakly-Convex Cases}: Common methods applied to analyze convergence of the proposed algorithm with respect to convex objective functions are no longer effective for non-convex objective functions. This stems from the observation that the typical stationarity measures $\mathbf{x}_t-\mathbf{x}^*$ and $f(\mathbf{x}_t)-f(\mathbf{x}^*)$ deteriorate and cannot precisely reflects the continuous evolution of the algorithm. Since weakly convex problems natuarally admit continuous measure of stationarity through implicit smoothing, we introduce the Moreau envelope \cite{1970ConvexAnalysis} which is a key construction for the problem. In this article, theoretical results are provided that the proposed algorithm generates an iterate sequence for each agent that subsequentially converges to a stationary point identified by the Moreau envelope. In addition, convergence rates are derived associated with quantization error $\Delta$, number of agents $N$, Lipchitz parameter $L$, and two stepsizes $\alpha_k$ and $\beta_k$.

\subsection{Related Works}
We briefly review the related works on distributed consensus from two different perspectives, i.e., imperfect information exchange for network consensus and convergence results for non-convex problems. In addition to quantized information, delayed information caused by communication latency is another kind of imperfect information. Methods for delayed information have been widely discussed in \cite{2021CompensatingActuator,2021ConsensofContinuous,2021DistributedAsynchronous,2022RobustCooperative,2022ConsensusStability}. In these works, the evolution of the algorithm is viewed as (stochastic) differential equations and the delay is transformed into the corresponding delayed timesteps and agent status. 

Convergence analysis of distributed gradient methods with random quantization is a novel area where the pioneering work is conducted by \cite{2019QuantizedDecentralized}. The authors consider decentralized consensus optimization for strongly convex settings. It is reported in this work that under standard strong convexity and smoothness assumptions for the objective function, quantization vanishes under customary conditions for quantizers. Based on a similar paradigm, following works \cite{2021ConvergenceRates,2021FastConvergenceRates} extend the result in \cite{2019QuantizedDecentralized} from two distinct viewpoints. To be specific, theoretical results on upper bounds of the convergence rates for both convex and strongly convex objective functions are provided in \cite{2021ConvergenceRates}, which extends the convergence results of strongly convex settings to convex settings. Established from previous results \cite{2019QuantizedDecentralized,2021ConvergenceRates}, it is evident that quantization slows convergence speed of the distributed gradient methods compared with ideal cases. To settle the problem, an adaptive quantization method is proposed \cite{2021FastConvergenceRates} which applies a time-varying quantization scheme with diminishing quantization errors. The authors show that the proposed scheme achieves the same convergence rate as that for perfect communications. Despite a similar quantization scheme, our work distinguishes the aforementioned results with respect to two aspects. First, general conclusions are derived on the convergence of weighted time-average parameters. Moreover, we extend convergence results from convex settings to weakly convex settings. 

Most recent works on analysis of convergence for non-convex problems focus on weakly convex settings \cite{2021DistributedStochasticConsensus,2022DecentralizedNonConvex,2022UnifiedRefinedConvergence}. In \cite{2021DistributedStochasticConsensus}, a distributed algorithm with Nesterov momentum for accelerated optimization of non-convex and non-smooth problems is proposed, and achieves $\mathcal{O}(1/\varepsilon^2)$ computaion complexity. Via a deep investigation into typical points of methods for non-convex problems, authors \cite{2022UnifiedRefinedConvergence} propose a general stochastic unified decentralized algorithm and establish the convergence under specific settings. From a different aspect, the problem of non-convex learning with linearly coupled constraints is studied \cite{2022DecentralizedNonConvex}. Except for weakly convex settings, there is a pioneering work on weakly quasi-convex settings \cite{2021Counterpart}, which improves non-asymptotic bounds in the convex setting for stochastic gradient descent (SGD) and applies to a class of weakly quasi-convex problems. 
\section{Preliminaries}
\label{sec:model}

\subsection{Notations and Definitions}
We first introduce some notations and definitions which will be used throughout the article. For a vector $\mathbf{x}$ and a matrix $\mathbf{X}$, $\|\mathbf{x}\|$ and $\|\mathbf{X}\|$ are used to represent the Euclidean norm and the Frobenius norm, respectively. We use $\langle \mathbf{x},\mathbf{y} \rangle$ to represent the inner product $\mathbf{x}^T\mathbf{y}$ of two vectors $\mathbf{x}$ and $\mathbf{y}$. Let $\mathbf{1}$ be the vector with all entries $1$ and $\mathbf{I}$ be the identity matrix. The mathematical expectation is expressed by $\mathbb{E}[\cdot]$. Given a closed convex set $\mathcal{X}$, the projection of $\mathbf{x}$ onto $\mathcal{X}$ is denoted by $\mathcal{P}_{\mathcal{X}}[\mathbf{x}]$. For a sequence of vectors $\{\mathbf{x}_1, \mathbf{x}_2, \dots, \mathbf{x}_N\}$ in $\mathbb{R}^d$, $\bar{\mathbf{x}}$ is used to represent the mean vector $\bar{\mathbf{x}} = \frac{1}{N} \sum_{i=1}^N \mathbf{x}_i$, and $\mathbf{X} \in \mathbb{R}^{n\times d}$ stands for the corresponding matrix comprised by the sequence. Throughout the article, a sequence $\{\mathbf{x}_n\}$ is said to converge to some $\mathbf{x}^*$ with convergence rate $\chi$ if 
\begin{equation}
  \|\mathbf{x}_n - \mathbf{x}^*\|^2 \in \mathcal{O}\left(\chi\right).
\end{equation}
In addition, we introduce the following necessary definitions to be considered in the context.

\begin{definition}[$\mu$-Strongly Convex]
  \label{def:strong-convexity}
  $f$ is strongly convex with parameter $\mu$, if $\forall \mathbf{x}, \mathbf{x}' \in \mathbb{R}^d$,
  \begin{equation}
    f(\mathbf{x}') \ge f(\mathbf{x}) + \langle \mathbf{g}(\mathbf{x}), \mathbf{x}'-\mathbf{x}\rangle + \frac{\mu}{2} \|\mathbf{x}'-\mathbf{x}\|^2,
  \end{equation}
  where $\mathbf{g}(\mathbf{x})$ denotes the subgradient of $f$ estimated at $\mathbf{x}$. 
\end{definition}

\begin{definition}[$\rho$-Weakly Convex]
  \label{def:weak-convexity}
  $f$ is weakly convex with parameter $\rho$, if $\forall \mathbf{x}, \mathbf{x}' \in \mathbb{R}^d$,
  \begin{equation}
    f(\mathbf{x}') \ge f(\mathbf{x}) + \langle \mathbf{g}(\mathbf{x}), \mathbf{x}'-\mathbf{x}\rangle - \frac{\rho}{2} \|\mathbf{x}'-\mathbf{x}\|^2,
  \end{equation}
  where $\mathbf{g}(\mathbf{x})$ denotes the subgradient of $f$ estimated at $\mathbf{x}$. 
\end{definition}

In Definition \ref{def:strong-convexity}, the concept of ``subgradient'' is introduced to cope with the case where $f$ is non-smooth and the existence of the gradient is unavailable. The subgradient $\mathbf{g}$ of $f$ is defined on a subdifferential estimation set, i.e.,
\begin{equation}
  \partial f(\mathbf{x}) \triangleq \{ \mathbf{g}\in \mathbb{R}^d | f(\mathbf{y}) \ge f(\mathbf{x}) + \langle \mathbf{g}, \mathbf{y}-\mathbf{x} \rangle, \forall \mathbf{y}\in \mathbb{R}^d \},
\end{equation}
where $f$ is convex. If $f$ is $\rho$-weakly convex, there exists a convex function $h(x)$ such that $f(\mathbf{x}) = h(\mathbf{x}) - \frac{\rho}{2}\|\mathbf{x}\|^2$. Therefore, the subdifferential of $f$ can be written as $\partial f(\mathbf{x}) = \partial h(\mathbf{x}) - \rho \mathbf{x}$, and the corresponding subgradient $\mathbf{u} \in \partial f(\mathbf{x})$.

\subsection{Distributed Subgradient with Random Quantization}
We consider a network of agents that can exchange information locally, which is defined through a connected and undirected graph $\mathcal{G} = (\mathcal{V},\mathcal{E})$, where $\mathcal{V} = \{1,2,\dots,N\}$ denotes the set of nodes and $\mathcal{E} \subset \mathcal{V}\times \mathcal{V}$ represents the set of links among the nodes. Each agent is associated with a non-smooth function $f_i: \mathcal{X} \rightarrow \mathbb{R}$ defined on a compact set $\mathcal{X} \subset \mathbb{R}^d$ which is known to all agents. The objective is to optimize an averaging function defined as
\begin{equation}
  \label{eqn:optimization-obj}
  \min_{\mathbf{x}\in\mathcal{X}} f(\mathbf{x}) = \frac{1}{N} \sum_{i=1}^N f_i(\mathbf{x}),
\end{equation}
where each $f_i$ is only available to agent $i$. A distributed consensus subgradient method is applied where each agent $i$ maintains a local variable $\mathbf{x}_i\in \mathbb{R}^d$. To make all $\mathbf{x}_i$'s converge to the optimal solution to (\ref{eqn:optimization-obj}), it is necessary for each agent $i$ to communicate with neighboring agents to exchange information about $\mathbf{x}_i$. We denote by $\mathcal{N}_i = \{j\in \mathcal{V}: (i,j)\in\mathcal{E}\}\cup \{i\}$ the set of agent $i$'s neighbors including agent $i$ itself. To capture the topology of $\mathcal{G}$, an adjacency matrix $\mathbf{A} = [a_{ij}]_{N\times N}$ is used to demonstrate the connectivity between agents.

To derive theoretical guarantees for convergence results of solutions to (\ref{eqn:optimization-obj}), several common assumptions are adopted in the context as follows: 

\begin{assumption}[Lipchitz]
  \label{asp:Lipchitz}
  The local functions $f_i, \forall i$ are $L_i$-Lipchitz continuous, i.e.,
  \begin{equation}
    \left| f_i(\mathbf{x}) - f_i(\mathbf{x}') \right| \le L_i \| \mathbf{x}-\mathbf{x}' \|, \quad \forall \mathbf{x},\mathbf{x}' \in \mathcal{X},
  \end{equation}
  where $\|\cdot\|$ represents the Euclidean norm of a vector. Furthermore, taking $L=\sum_{i=1}^N L_i$, we obtain
  \begin{equation}
    \left| f(\mathbf{x}) - f(\mathbf{x}') \right| \le L \| \mathbf{x}-\mathbf{x}' \|, \quad \forall \mathbf{x},\mathbf{x}' \in \mathcal{X}.
  \end{equation}
\end{assumption}

\begin{assumption}[Doubly Stochastic]
  \label{asp:doubly-stochastic}
  The matrix $\mathbf{A}$ is irreducible, symmetric and aperiodic. Furthermore, the matrix $\mathbf{A}$ is doubly stochastic, i.e., $\sum_{i=1}^N a_{ij}=\sum_{j=1}^N a_{ij} = 1$. The entries $a_{ij}$ are positive if and only if $(i,j) \in \mathcal{E}$.
\end{assumption}

Note that Assumption \ref{asp:doubly-stochastic} indicates that $\mathbf{A}$ has the largest eigenvalue $1$ and other eigenvalues strictly less than $1$. We denote the second largest eigenvalue of $\mathbf{A}$ by $\sigma_2$ which will be useful in the convergence analysis. 

In practice, information exchange between agents is affected by communication bandwidth, physical equipment limitations, and signal inteferences, which result in deteriorated feasible transmission data rates. Lower data rates will cause higher transmission latency, which potentially harms performances of distributed optimization. An alternative solution to the problem is quantization. By quantizing the data to be transmitted, transmission overhead can be significantly reduced. 

To be specific, there exists some integer $\ell$ for any $x \in \mathbb{R}$ satisfying $\ell \le x < \ell+1$. For a $b$-bit quantization scheme, $x$ falls in an interval $[\tau_i, \tau_{i+1})$, where $\tau_i = \ell + 2^{-b}(i-1)$. In this article, we consider a stochastic $b$-bit quantization scheme based on relative location in the interval. Denoted by $q$, $x$ is represented by $\tau_i$ or $\tau_{i+1}$ randomly according to
\begin{equation}
  \label{eqn:quantization}
  \mathrm{P}(q=\tau_{i+1}) = 2^b(x - \tau_i), \ \mathrm{P}(q=\tau_i) = 2^b(\tau_{i+1} - x),
\end{equation}
which indicates that $\mathbb{E}[q] = x$, $\var[q]\le 4^{-b-1}$. Without loss of generality, we use $\Delta$ to stand for the length of quantization interval in the context, i.e., $2^{-b}$ in this case. 

\begin{algorithm}[t]
  \SetAlgoLined
  \textbf{Initialization}: Each agent $i$ initializes a local parameter $\mathbf{x}_i(0) \in \mathcal{X}$, shares common stepsizes $\{\alpha_k,\beta_k\}$, and determines $\{\gamma_k^t\}$\;
  \For{$k \leftarrow 1$ \KwTo $K$}{
    Each node $i$ sends quantized vector $\mathbf{q}_i(k)$ to neighboring nodes $j \in \mathcal{N}_i$\;
    Each node $i$ receives $\mathbf{q}_j(k)$ from node $j \in \mathcal{N}_i$\;
    Each node updates local parameter $\mathbf{x}_i(k)$ as (\ref{eqn:xi-update})\;
    Update the time-average $\mathbf{z}_i(k)$ of $\{\mathbf{x}_i(k)\}$ as (\ref{eqn:zi-update}).
  }
  \caption{Generalized Distributed Subgradient Method with Random Quantization (GDSRQ).}
  \label{alg:GDSRQ}
\end{algorithm}

Next, we proceed to introduce the algorithm procedure for generalized distributed subgradient with random quantization (GDSRQ), which is an extension of the previous work \cite{2021ConvergenceRates}. The iteration update for GDSRQ can be written as
\begin{equation}
  \label{eqn:xi-update}
  \mathbf{x}_{i,k+1} = \mathcal{P}_{\mathcal{X}} [\mathbf{v}_{i,k} - \alpha_k \mathbf{g}_{i,k}],
\end{equation}
where $\mathbf{v}_{i,k} = (1-\beta_k) \mathbf{x}_{i,k} + \beta_k \sum_{j \in \mathcal{N}_i} a_{ij} \mathbf{q}_{j,k}$, $\mathbf{q}_{j,k}$ is the randomly quantized data vector of $\mathbf{x}_{j,k}$ according to (\ref{eqn:quantization}), $\mathbf{g}_{i,k}$ is the subgradient of $f_i$ evaluated at $\mathbf{x}_{i,k}$, and $\alpha_k, \beta_k$ are two time-scale stepsizes. We note that the iteration update of GDSRQ is the same as that in \cite{2021ConvergenceRates}, while the output of the algorithm is generalized as follows. Except for the local parameter $\{\mathbf{x}_{i,k}\}$, a weighted time-average of $\{\mathbf{x}_{i,k}\}$ denoted by $\{\mathbf{z}_{i,k}\}$ is also maintained as
\begin{equation}
  \label{eqn:zi-update}
  \mathbf{z}_{i,k} = \sum_{t=0}^k \gamma_k^t  \mathbf{x}_{i,t},
\end{equation}
where $\{\gamma_k^t\}$ are time-average weights independent of $\{\alpha_k,\beta_k\}$. We will show later that $\{\mathbf{z}_{i,k}\}$ provides a lower-bound approximation to the expectation of $f(\mathbf{x}_i)$, which demonstrates the convergence process towards the optimal value $f^*$. The complete procedure of GDSRQ is presented as Algorithm \ref{alg:GDSRQ}.

\section{Main Results and Convergence Analysis}
In this section, we state the main convergence results of GDSRQ. We consider two main cases where the objective functions are (strongly) convex and weakly convex, respectively. For (strongly) convex objective functions, the convergence analysis follow the standard outline for a general distributed stochastic subgradient method. In comparison, such methodology is not applicable for weakly convex objective functions since underlying prerequisites of common approaches for convex cases are violated. To resolve this problem, the Moreau envelope $\varphi_{1/\bar{\rho}}(\cdot)$ is introduced to provide a continuous measure to monitor the progress of GDSRQ \cite{2019WeaklyConvex}. By deriving the convergence properties of $\varphi_{1/\bar{\rho}}(\cdot)$ and the gradient $\partial \varphi_{1/\bar{\rho}}(\cdot)$, we provide the convergence result of GDSRQ when the objective functions are weakly convex. 

\subsection{Analysis of Convex Objective Functions}
Throughout this subsection, we use $\mathbf{X}_k$, $\mathbf{Q}_k$, and $\mathbf{Y}_k$ to represent the matrix at iteration $k$ comprised by $\{\mathbf{x}_{i,k}\}$, $\{\mathbf{q}_{i,k}\}$ and $\{\mathbf{x}_{i,k} - \bar{\mathbf{x}}_k\}$, respectively. To be specific, $\mathbf{X}_k$ is expressed by $[\mathbf{x}_1(k), \mathbf{x}_2(k), \cdots, \mathbf{x}_N(k)]^T \in \mathbb{R}^{N\times d}$. Moreover, $r_k$ is used to denote the deviation of $\bar{\mathbf{x}}_k$ from the minimizer $\mathbf{x}^*$ of the optimization objective function (\ref{eqn:optimization-obj}), i.e., $r_k=\|\bar{\mathbf{x}}_k - \mathbf{x}^*\|^2$.

We first state without proof the following lemmas to facilitate the theoretical derivation in the context. For more detailed analysis on the lemmas, it directly refers to \cite{2021ConvergenceRates}.
\begin{lemma}
  \label{lma:x-deviation}
  Let $\mathscr{F}_k$ be the filtration containing all the history upto time $k$, i.e., $\mathscr{F}_k = \{\mathbf{X}_0, \mathbf{Q}_0, \mathbf{X}_1, \mathbf{Q}_1, \dots, \mathbf{X}_k, \mathbf{Q}_k\}$. In terms of the generated sequences $\{x_{i,k}\}, \forall i\in \mathcal{V}$, we have 
  \begin{equation}
    \label{eqn:Yk-recursive}
    \begin{aligned}
      \mathbb{E}\left[ \|\mathbf{Y}_{k+1}\|^2| \mathscr{F}_k \right] &\le \left[ 1-(1-\sigma_2)\beta_k \right] \|\mathbf{Y}_k \|^2 \\
      &+ N\sigma_2^2 \Delta^2 \beta_k^2 + \frac{4\alpha_k^2 L^2 (\beta_0+1)}{\beta_k (1-\sigma_2)},
    \end{aligned}
  \end{equation}
  where $\sigma_2$ is the second largest singular value of adjacency matrix $\mathbf{A}$. If $\sum_{k=1}^\infty \beta_k =\infty$, $\sum_{k=1}^\infty \frac{\alpha_k^2}{\beta_k}<\infty$ and $\sum_{k=1}^\infty \beta_k^2 <\infty$, we have
  \begin{equation}
    \sum_{k=1}^\infty \beta_k \|\mathbf{Y}_k\|^2 < \infty \Longrightarrow \lim_{k\rightarrow \infty} \mathbb{E} \|\mathbf{Y}_k\| = 0.
  \end{equation}
\end{lemma}

\begin{lemma}
  \label{lma:x-avg-error}
  Let $\mathbf{x}^*$ be a minimizer of $f(\mathbf{x})$, then we have
  \begin{equation}
    \begin{aligned}
      \mathbb{E}\left[ r_{k+1}|\mathscr{F}_k \right] \le r_k + \frac{6L^2 \alpha_k^2}{N(1-\beta_0)} + \frac{2L^2\alpha_k^2}{N\beta_k} + \Delta^2 \beta_k^2 \\
      + \frac{2\beta_k \|\mathbf{Y}_k\|^2}{N} - \frac{2\alpha_k}{N} \sum_{i=1}^N \mathbf{g}_i^T(\mathbf{x}_{i,k}) (\mathbf{x}_{i,k} - \mathbf{x}^*),
    \end{aligned}
  \end{equation}
  Furthermore, if $\sum_{k=1}^\infty \beta_k =\infty$, $\sum_{k=1}^\infty \alpha_k =\infty$, $\sum_{k=1}^\infty \frac{\alpha_k^2}{\beta_k} <\infty$ and $\sum_{k=1}^\infty \beta_k^2 <\infty$, the limit of $\mathbb{E}\left[ r_{k}\right]$ exists and satisfies
  \begin{equation}
    \label{eqn:rk-limit}
    \lim_{k\rightarrow \infty} \mathbb{E}\left[ r_{k}\right] = 0.
  \end{equation}
\end{lemma}

As can be conlcuded from Lemma \ref{lma:x-deviation} and Lemma \ref{lma:x-avg-error}, $\mathbf{x}_{i,k}$ converges to the optimal parameter vector $\mathbf{x}^*$ for each node $i$ as $k \rightarrow \infty$. Since the objective function $f$ is assumed to be Lipchitz continuous, the convergence process of $\mathbf{x}_{i,k}$ provides an lower bound guarantee for the convergence speed of $f$. Nevertheless, it is still difficult to explicitly express the convergence speed of $f$. To clearly demonstrate the convergence process of $f$, we use a weighted time-average of $\{\mathbf{x}_{i,k}\}$ to capture the characteristics of evolution of $f$ with respect to step sizes and quantization bits. 
 
% \begin{lemma}
%   \label{lma:estimation-error}
%   Let $\mathbf{x}^*$ be a minimizer of $f(\mathbf{x})$, then we have
%   \begin{equation}
%     \lim_{k\rightarrow \infty} \mathbf{x}_{i,k} = \mathbf{x}^*, \quad \text{a.s.} \quad \forall i\in \mathcal{V},
%   \end{equation}
%   which indicates that $\lim_{k\rightarrow \infty} r_k = 0$.
% \end{lemma}

Before proceeding to provide the main theorem, we introduce the following lemma to facilitate the derivation:

\begin{lemma}
  \label{lma:convergence-of-production}
  Let $0<\delta\le 1$ be a positive constant, and $\{x_k\}$ be a positive decreasing sequence defined on $\mathbb{R}$ with $0<x_k \le 1, \forall k$. Assuming the series of $\{x_k\}$ is divergent while quadratically convergent, we have $\prod_{k=0}^\infty (1-\delta x_k) = 0$. Furthermore, if we assume $\lim_{k\rightarrow \infty} x_k(k+1) = \infty$, then $\sum_{k=0}^\infty \prod_{t=0}^k (1-\delta x_k) < \infty$.
\end{lemma}

We provide sufficient conditions that $\{\gamma_k^t\}$ should satisfy for the convergence of $\{\mathbf{z}_{i,k}\}$ towards the minimizer of (\ref{eqn:optimization-obj}). 

% #################### Main Theorem #######################
\begin{theorem}
  \label{thm:strongly-convex-convergence}
  Let $\mathbf{x}^*$ be a minimizer of \textbf{strongly convex} function $f(\mathbf{x})$, and let decreasing stepsizes $\{\alpha_k\}$, $\{\beta_k\}$ satisfy
  \begin{subequations}
    \begin{equation}
      \label{eqn:assumption-stepsizes1}
      \lim_{k\rightarrow\infty} k\beta_k=\infty, \lim_{k\rightarrow\infty} \frac{1}{k\alpha_k}\le C, \sum_{k=0}^\infty \beta_k^2 < \infty, \ \sum_{k=0}^\infty \frac{\alpha_k^2}{\beta_k} < \infty,
    \end{equation}
    \begin{equation}
      \label{eqn:assumption-stepsizes2}
      \sum_{k=0}^\infty \alpha_k = \infty, \ \sum_{k=0}^\infty \beta_k = \infty, 
    \end{equation}
  \end{subequations}
  where $C$ is chosen to be a positive constant. In addition, suppose that each node $i$ maintains a local random variable $\mathbf{z}_{i,k}$ initialized arbitrarily and updated as
  \begin{equation}
    \label{eqn:local-estimate}
    \mathbf{z}_{i,k} = \sum_{t=0}^k \gamma_k^t \mathbf{x}_{i,k},
  \end{equation}
  where $0<\gamma_k^t<1$ is designed satisfying $\forall j \le k$, $\forall k$,
  \begin{subequations}
    \begin{equation}
      \label{eqn:gamma-ini}
      \lim_{k\rightarrow \infty}\sum_{t=0}^k \frac{\gamma_k^t}{\alpha_t} \beta_t^2 =0, \ \lim_{k\rightarrow \infty} \sum_{t=0}^k \frac{\gamma_k^t \alpha_t}{\beta_t} = 0, \ \sum_{t=0}^k \gamma_k^t=1,
    \end{equation}
    \begin{equation}
      \label{eqn:gamma-constraint}
      \frac{\gamma_k^{j+1}}{\alpha_{j+1}} (1-\mu \alpha_{j+1}) \le \frac{\gamma_k^j}{\alpha_j}, \ \lim_{k\rightarrow \infty} \sum_{j=0}^{k-1} \left|\frac{\gamma_k^j}{\alpha_j} - \frac{\gamma_k^{j+1}}{\alpha_{j+1}}\right| = 0.
    \end{equation}
  \end{subequations}
  Then for all $i \in \mathcal{V}$ we have
  \begin{equation}
    \lim_{k\rightarrow \infty} \mathbf{z}_{i,k} = \mathbf{x}^*, \quad \text{a.s.}
  \end{equation}
\end{theorem}
% #################### Convex #######################

For Theorem \ref{thm:strongly-convex-convergence} builds on the condition that $f$ is strongly convex, the proof is established on basic methods of stochastic gradient analysis. Note that the fundamental methodology of stochastic gradient analysis naturally applies to convex objective functions. Therefore, by relaxing condition (\ref{eqn:gamma-ini}) or (\ref{eqn:gamma-constraint}), we can intuitively generalize the conclusion in Theorem \ref{thm:strongly-convex-convergence} to the case where $f$ is convex as follows: 

\begin{proposition}
  \label{prop:convex-convergence}
  Let $\mathbf{x}^*$ be a minimizer of \textbf{convex} function $f(\mathbf{x})$, and let decreasing stepsizes $\{\alpha_k\}$, $\{\beta_k\}$ satisfy (\ref{eqn:assumption-stepsizes1}) and (\ref{eqn:assumption-stepsizes2}). Suppose that $\mathbf{z}_{i,k}$ is updated as (\ref{eqn:local-estimate}), and $0<\gamma_k^t<1$ is designed satisfying (\ref{eqn:gamma-ini}) for all $j\le k$, $\forall k$ and
  \begin{equation}
    \frac{\gamma_k^{j+1}}{\alpha_{j+1}} \le \frac{\gamma_k^j}{\alpha_j}, \ \lim_{k\rightarrow \infty} \left(\frac{\gamma_k^0}{\alpha_0} - \frac{\gamma_k^k}{\alpha_k} \right)= 0.
  \end{equation}
  Then for all $i \in \mathcal{V}$ we have
  \begin{equation}
    \lim_{k\rightarrow \infty} \mathbf{z}_{i,k} = \mathbf{x}^*, \quad \text{a.s.}
  \end{equation}
\end{proposition}

Theorem \ref{thm:strongly-convex-convergence} and Proposition \ref{prop:convex-convergence} reveal the sufficient conditions for the convergence of $\{\mathbf{z}_{i,k}\}$ to $\mathbf{x}^*$. To explicitly express the convergence rates of $f(\mathbf{z}_{i,k})$, exponentially diminishing stepsizes are investigated as a typical example to provide more specific insights into the factors which exert significant influences on the convergence of $f$. 

\begin{proposition}
  \label{prop:convergence-rates-special-case}
  Let $\mathbf{x}^*$ be a minimizer of \textbf{convex} function $f(\mathbf{x})$, and let decreasing stepsizes $\{\alpha_k\}$, $\{\beta_k\}$ be set as 
  \begin{equation}
    \alpha_k = \alpha_0 (k+1)^{-\lambda_\alpha}, \ \beta_k = \beta_0 (k+1)^{-\lambda_\beta},
  \end{equation}
  where $1/2 < \lambda_\alpha\le 1$, $1/2 < \lambda_\beta < 1$, $2\lambda_\beta > \lambda_\alpha$ and $(1-\sigma_2)\beta_0<1$. Suppose that $\mathbf{z}_{i,k}$ is updated as (\ref{eqn:local-estimate}), and $0<\gamma_k^t<1$ is designed satisfying
  \begin{equation}
    \gamma_k^t = \frac{(t+1)^{-\lambda_\gamma}}{\sum_{t=0}^k (t+1)^{-\lambda_\gamma}},
  \end{equation}  
  where $\lambda_\alpha \le \lambda_\gamma \le 1$. Then for all $i \in \mathcal{V}$ we obtain
  \begin{equation}
    \label{eqn:f-convergence-speed}
    \begin{aligned}
      &\mathbb{E}[f(\mathbf{z}_{i,k})] - f^* \le \\
      &\mathcal{O}\left( \frac{L^2}{N} (k+1)^{\lambda_\beta - \lambda_\alpha} + \sigma_2^2 \Delta^2 (k+1)^{1+\lambda_\alpha - 4\lambda_\beta} \right).
    \end{aligned}
  \end{equation}
  Therefore, the convergence rate order $\chi$ of $\mathbf{z}_{i,k}$ is
  \begin{equation}
    \chi = \max\{\lambda_\beta - \lambda_\alpha, 1+\lambda_\alpha-4\lambda_\beta\}.
  \end{equation}
  Moreover, the optimal convergence rate is obtained as $\frac{1}{2}-\frac{3}{2}\lambda_\beta$ by letting $\lambda_\alpha = \frac{5}{2}\lambda_\beta - \frac{1}{2}$ if $\lambda_\beta \le \frac{3}{5}$.
\end{proposition}

As can be deduced from (\ref{eqn:f-convergence-speed}), the convergence rate is independent of the order of $\gamma_k^t$ when the weight and step sizes are exponentially diminishing. Moreover, the gap between current function value and the optimal value increases with respect to the length of quantization intervals $\Delta$, the second largest eigenvalue $\sigma_2$ and Lipchitz parameter $L$. The reasons for the result are three-fold. First, lower bits of quantization produce larger length of quantization intervals, cause more quantization error, and deteriorate training performances. In addition, $1-\sigma_2$ represents the spectral gap of a Markov chain corresponding to the doubly stochastic matrix, which is a quantitative measure of ergodicity of the Markov chain. Therefore, the greater $1-\sigma_2$ is, the faster is the speed of convergence of any initial distribution to an equilibrium. Moreover, $L$ provides information about ``sharpness'' of the objective function. Given some fixed parameter improvement, a larger $L$ means less objective function value drop. 

% #################### Weakly Convex #######################

\subsection{Weakly-Convex Case}
In this subsection, we discuss the non-convex objective functions. That is, local functions $f_i(\mathbf{x})$ and the corresponding global function $f(\mathbf{x})$ are all weakly convex with parameter $\rho$. The derivation of non-convex functions follows a different methodology. Let $\varphi(\mathbf{x}) = f(\mathbf{x}) + \mathrm{1}_{\mathcal{X}}(\mathbf{x})$, where $\mathrm{1}_{\mathcal{X}}(\mathbf{x})$ is the indicator function of $\mathcal{X}$. The Moreau envelope is defined as \cite{1970ConvexAnalysis}
\begin{equation}
  \label{eqn:Moreau-envelope}
  \varphi_\theta (\mathbf{x}) := \min_{\mathbf{y}} \varphi(\mathbf{y}) + \frac{1}{2\theta} \|\mathbf{y} - \mathbf{x}\|^2, \quad 0<\theta<1/\rho.
\end{equation}
As can be deduced from (\ref{eqn:Moreau-envelope}), $\varphi(\mathbf{y})+\frac{1}{2\theta}\|\mathbf{y} - \mathbf{x}\|^2$ is strongly convex of parameter $\frac{1}{\theta}-\rho$ with respect to $\mathbf{y}$. Therefore, there exists a unique minimizer to the optimization problem in (\ref{eqn:Moreau-envelope}). That is, the minimizer is written as
\begin{equation}
  \hat{\mathbf{x}} := \argmin_{\mathbf{y}} \varphi(\mathbf{y}) + \frac{1}{2\theta} \|\mathbf{y} - \mathbf{x}\|^2,
\end{equation}
which is called the proximal mapping written as $\hat{\mathbf{x}} = \text{prox}_{\theta\varphi}(\mathbf{x})$. We note that the proximal mapping is only used in our analysis and will not be included in the algorithm. Our aim is to provide the convergence result that the Moreau envelope sequence $\varphi_\theta(\bar{\mathbf{x}}_t)$ converges to the unique minimum value with a geometrically diminishing step size. 

Apart from the two assumptions mentioned earlier, we make an auxiliary assumption on the boundedness of $f$ as follows:
\begin{assumption}
  \label{asp:lower-bounded}  
  The objective function $f$ is lower bounded, i.e., there exists some $f^\dagger$ such that $f(\mathbf{x}) \ge f^\dagger$ for $\forall \mathbf{x}$. Moreover, the subgradient of $f_i$ is Lipchitz continuous with parameter $\nu$, i.e., $\|\mathbf{g}_i(\mathbf{x})-\mathbf{g}_i(\mathbf{y})\|\le \nu\|\mathbf{x} - \mathbf{y}\|$. 
\end{assumption}

% We substitute $\mathbf{u}_{i,k}$ into the gradient term in (\ref{eqn:xi-update}) which is the subgradient of $f(\mathbf{v}_{i,k})$, i.e., $\{\mathbf{x}_{i,k+1}\}$ updates according to 
% \begin{equation}
%   \label{eqn:xi-update-new}
%   \mathbf{x}_{i,k+1} = \mathcal{P}_{\mathcal{X}} [\mathbf{v}_{i,k} - \alpha_k \mathbf{u}_{i,k}].
% \end{equation}
A significant element of convergence analysis is the deviation of individual errors $\|\mathbf{Y}_{k}\|$. In the former subsection, the convergence of $\|\mathbf{Y}_{k}\|$ was established for convex objective functions. For weakly convex problems, such result still holds based on the proof of Lemma \ref{lma:x-deviation} reported in \cite{2021ConvergenceRates}. In addition, we state without proof a fundamental property of weakly convex functions which is analogous to the convex counterpart: 

\begin{lemma}[\cite{2022WeaklyConvex}]
  \label{lma:weakly-convex-property}
  If $f(\mathbf{x})$ is $\rho$-weakly convex, it follows that
  \begin{equation}
    f\left( \sum_{i=1}^N c_i \mathbf{x}_i \right) \le \sum_{i=1}^N c_i f(\mathbf{x}_i) + \frac{\rho}{2} \sum_{i=1}^{N-1} \sum_{j=i+1}^N c_i c_j \|\mathbf{x}_i - \mathbf{x}_j\|^2,
  \end{equation}
  where $c_i \ge 0, \forall i$ and $\sum_{i=1}^N c_i =1$.
\end{lemma}

Moreover, we provide an important lemma on the Lipchitz continuity of proximal mapping, which is very useful to proof of the main theorem in this subsection. 

\begin{lemma}
  \label{lma:weakly-convex-Lipchitz}
  If $\varphi(\mathbf{x})$ is $\rho$-weakly convex, the proximal mapping with parameter $0 < \theta < 1/\rho$ satisfies $\forall \mathbf{x},\mathbf{y}$
  \begin{equation}
    \|\text{prox}_{\theta \varphi}(\mathbf{x}) - \text{prox}_{\theta \varphi}(\mathbf{y})\| \le \frac{1}{1-\theta \rho} \|\mathbf{x} - \mathbf{y}\|,
  \end{equation}
  where the proximal mapping $\text{prox}_{\theta \varphi}(\mathbf{x})$ is the counterpart of $\varphi_\theta(\mathbf{x})$, and $1-\theta \rho > 0$ ensures the inequality is well-defined. 
\end{lemma}

Based on the lemmas, we provide the convergence results of Algorithm \ref{alg:GDSRQ} for weakly convex objective functions as follows:

\begin{theorem}
  \label{thm:weakly-convex-convergence}
  Let $0<\theta < \frac{2}{3\rho}$ and $\{\mathbf{x}_{i,k}\}$ be generated by Algorithm \ref{alg:GDSRQ}. If $\{\alpha_k\}$ and $\{\beta_k\}$ satisfy
  \begin{equation}
    \sum_{k=1}^\infty \alpha_k = \infty, \ \sum_{k=1}^\infty \beta_k=\infty, \ \sum_{k=1}^\infty \beta_k^2 < \infty, \ \sum_{k=1}^\infty \frac{\alpha_k^2}{\beta_k} < \infty,
  \end{equation}
  then there exists $\tilde{\varphi}_\theta = \min_{\mathbf{y}} \varphi_\theta (\mathbf{y})$ such that 
  \begin{equation}
    \label{eqn:Moreau-envelop-converge}
    \lim_{k\rightarrow \infty} \varphi_\theta (\mathbf{x}_{i,k}) = \lim_{k\rightarrow \infty} \varphi_\theta (\bar{\mathbf{x}}_{k}) =  \tilde{\varphi}_\theta.
  \end{equation}
  Moreover, we have the following convergence rate
  \begin{equation}
    \label{eqn:Moreau-envelope-convergence-rate}
    \inf_{0 \le t \le T} \mathbb{E}\|\nabla \varphi_\theta (\bar{\mathbf{x}}_t)\|^2 \le \mathcal{O}\left( \frac{\sum_{t=0}^T L^2\alpha_t^2 + b_t + q_t}{N \sum_{t=0}^T \alpha_t} \right),
  \end{equation}
  where $L$ denotes the Lipchitz parameter of $f$, $N$ is the number of agents, $\Delta$ represents the quantization error of transmitted parameters, $b_t = \mathcal{O}\left( \frac{\alpha_t^2}{\beta_t} + \beta_t^2 \right)$ and $q_t = \mathcal{O}(\beta_t \alpha_t \Delta)$. 
\end{theorem}

\textbf{Remark.} (\ref{eqn:Moreau-envelop-converge}) reveals the fact that the Moreau envelope function values evaluated at $\bar{\mathbf{x}}_k$ converge to the optimum and local parameters $\mathbf{x}_{i,k}$ reach distributed consensus. The convergence rate (\ref{eqn:Moreau-envelope-convergence-rate}) is the decentralized counterpart of the centralized algorithm \cite{2019WeaklyConvex}. By comparing the convergence rate with that of the centralized algorithm, we notice that there are extra terms $b_t$ and $q_t$, which are interpreted as additional cost for distributed consensus and quantization errors. 
\section{Simulation}
\label{sec:simulation}

In this section, we apply the proposed GDSRQ to solve the linear regression problems, which have been the most popular verification method in machine learning conducted over a networks of agents. In the problem, we aim to find the linear relationship between the training set and the correpsonding label (some real value). To be specific, given a training set $\mathcal{D} = \{(\mathbf{a}_i,b_i) \in \mathbb{R}^d \times \mathbb{R}\}$ for $i \in \mathcal{V}$, the objective of the network is to learn some parameter $\mathbf{x}$ which optimizes 
\begin{equation}
  \min_{\mathbf{x}\in \mathcal{X}} \frac{1}{N}\sum_{i=1}^N f_i(\mathbf{x};\mathbf{a}_i,b_i),
\end{equation}
where $\mathcal{X} \subset \mathbb{R}^d$, $d=10$, and $f_i$ denotes the local loss function defined over the dataset. We consider the performance of the proposed algorithm on a symmetric undirected connected graph of 150 nodes, i.e., $|\mathcal{V}|=N=150$. In the network, the nodes' coordinates are generated randomly following uniform distribution over a bounded plane. Two nodes are viewed as connected if the distance between them is less than a predefined threshold $r$, which is set as $r=0.5$ in the simulation. Given the generated graph, connectivity is checked. If the graph is not connected, we regenerate a graph following the same procedure. 

In the network, we adopt a lazy Metropolis matrix as the adjacency matrix $\mathbf{A}$ corresponding to the generated graph $\mathcal{G}$:
\begin{equation}
  \mathbf{A} = [a_{ij}] = \left\{
  \begin{aligned}
    \frac{1}{2\max\{|\mathcal{N}_i|,|\mathcal{N}_j|\}}, && (i,j)\in \mathcal{E},\\
    0, && (i,j) \notin \mathcal{E}, i\neq j, \\
    1 - \sum_{j\in \mathcal{N}_i} a_{ij}, && i=j.
  \end{aligned}
  \right.
\end{equation}
Note that it is straightforward to verify $\mathbf{A}$ to be doubly stochastic. We use Matlab to compute the optimal solution to the optimization problem and use the results as a reference to examine the performance of the algorithm. 

\subsection{Convex Case}

In the first simulation for convex objective functions, we consider quadratic loss functions defined as
\begin{equation}
  \min_{\mathbf{x}\in [-1,1]^d} \frac{1}{N}\sum_{i=1}^N (\mathbf{a}_i^T\mathbf{x}-b_i)^2.
\end{equation}
We use a synthetic dataset which is generated by uniformly sampling from $[0,1]$, i.e., $(\mathbf{a}_i,b_i) \in [0,1]^{10} \times [0,1]$. 

\begin{figure}[t]
  \centering
  \includegraphics[width=0.48\textwidth]{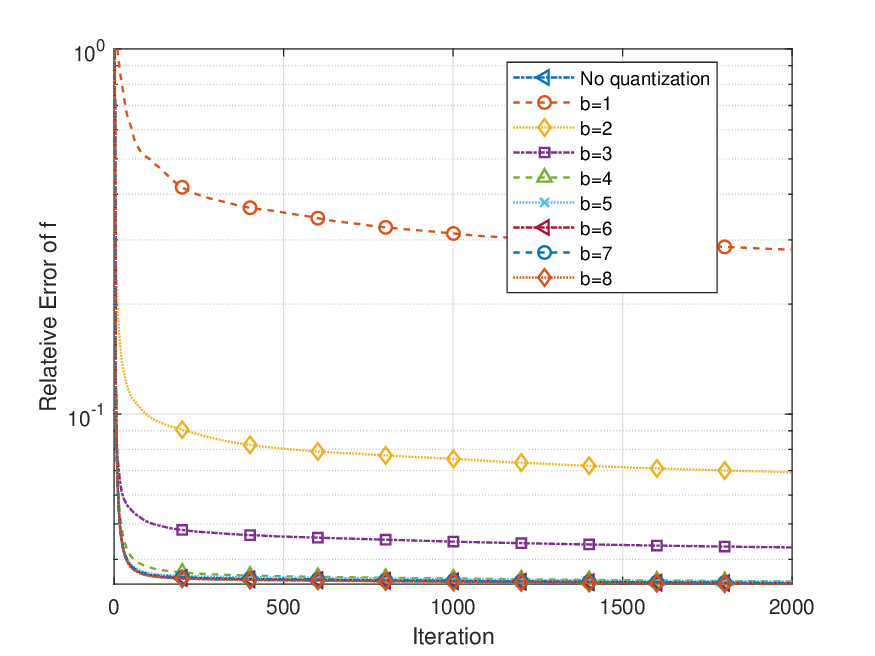}
  \caption{Convergence of function value under quadratic loss functions.}
  \label{fig:conv-f}
\end{figure}

\begin{figure}[t]
  \centering
  \includegraphics[width=0.48\textwidth]{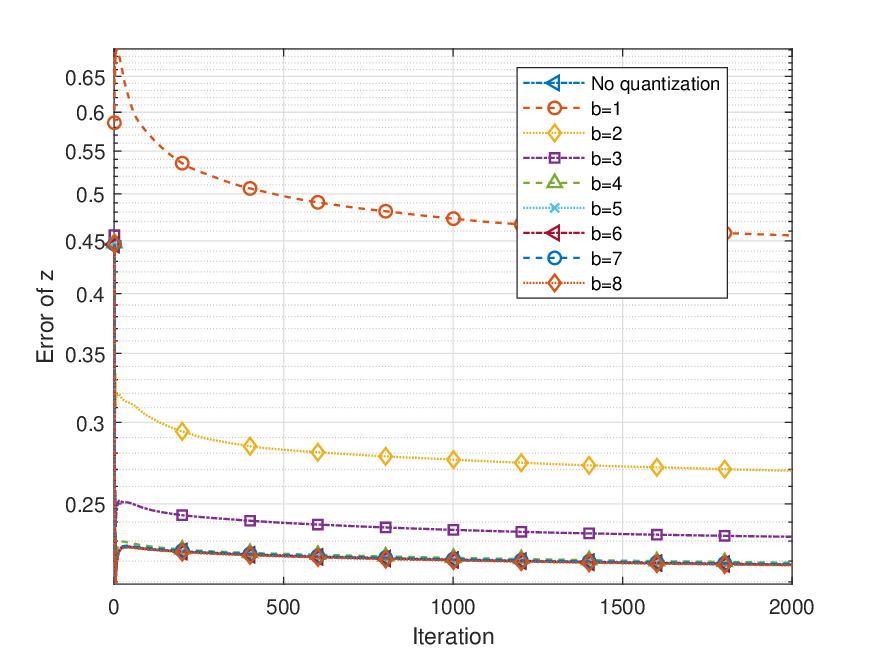}
  \caption{Convergence of $z$ under quadratic loss functions.}
  \label{fig:conv-z}
\end{figure}

In Fig. \ref{fig:conv-f}, we illustrate the results in Theorem \ref{thm:strongly-convex-convergence} by showing the convergence of square distances of the iterates to the optimal value. In the simulation, we chose $\alpha_k=1/k$ and $\beta_k = 1/k^{0.6}$, which satisfies the sufficient conditions for the convergence. The convergence of the time-average parameter $z$ is demonstrated in Fig. \ref{fig:conv-z}. As can be concluded from the figure, $z$ converges the optimal parameter correpsonding to the optimal global function value with respect to iteration $k$. Moreover, both figures exhibit the relationship between quantization bit and the convergence rates. As quantization bit increases, i.e., more accurate transmitted data, the convergence becomes faster. In addition, we can notice that the increase of quantization bit contribute much less when $b > 6$. This observation motivates us that there exists some threshold on the contribution of data accuracy. Without appropriate choices, the loss of increasing quantization bit will outweigh the gain. 

\subsection{Weakly Convex Case}

In the second experiment, a phase retrieval problem is considered as follows:
\begin{equation}
    \min_{\mathbf{x}\in \mathbb{R}^d} \frac{1}{NM}\sum_{i=1}^N \sum_{j=1}^M |(\mathbf{a}_{ij}^T\mathbf{x})^2-b_{ij}|,
\end{equation}
where the standard Gaussian measurements $a_{ij} \sim \mathcal{N}(0,\mathbf{I}_{d\times d})$ are independently and identically generated from a $d$-dimensional normal distribution. A target signal $\mathbf{x}^\dagger$ and an initial point $\mathbf{x}_0$ are generated uniformly on the unit sphere. $b_{ij}$ is set to be $(\mathbf{a}_{ij}^T \mathbf{x}^\dagger)^2$ for each $i,j$. 

\begin{figure}[t]
  \centering
  \includegraphics[width=0.48\textwidth]{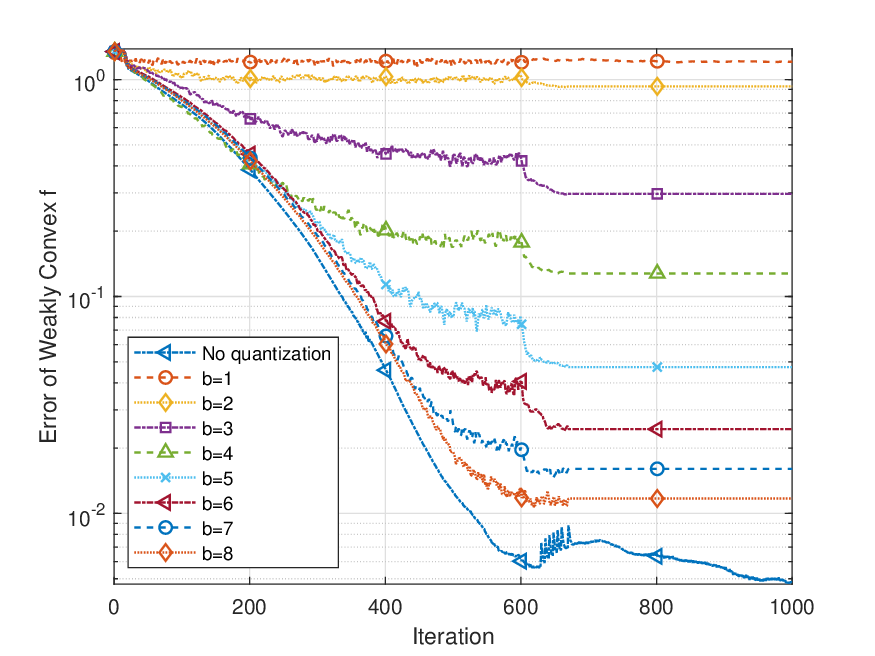}
  \caption{Convergence of function value for phase retrieval.}
  \label{fig:conv-weakly-f}
\end{figure}

\begin{figure}[t]
  \centering
  \includegraphics[width=0.48\textwidth]{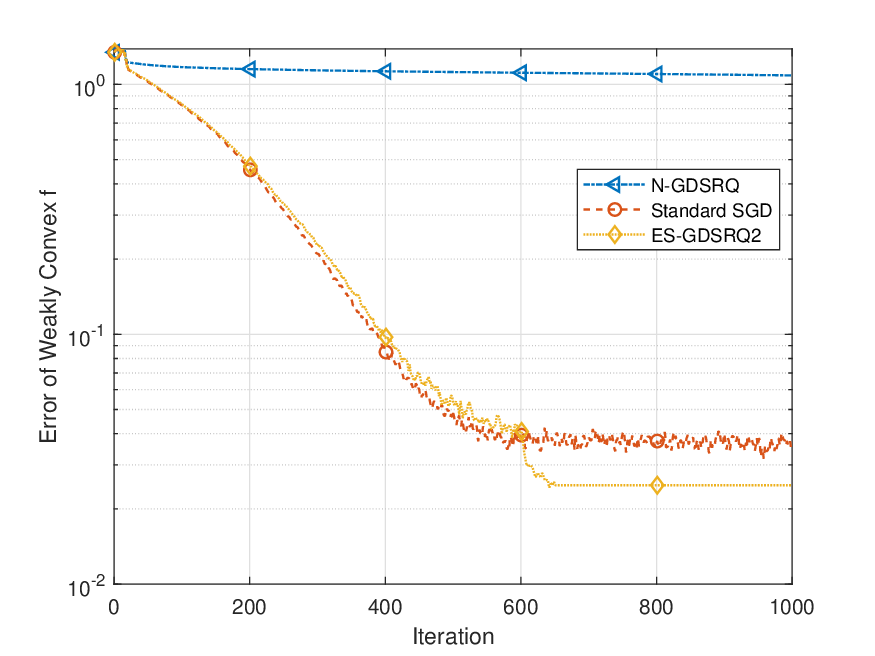}
  \caption{Comparisons of different methods. N-GDSRQ: Naive GDSRQ, Standard SGD: Standard distributed subgradient method, ES-GDSRQ2: Two-stage GDSRQ with early stopping.}
  \label{fig:comp}
\end{figure}

To accelerate the convergence of the proposed algorithm, we combine standard SGD method \cite{2022WeaklyConvex} and apply a popular technique in deep learning called early stopping \cite{2018NipsEralyStopping}. The accelerated algorithm is named as two-stage GDSRQ with early stoppping, which is distinguished from the naive version where $\beta_t = 1/t^{\frac{2}{3}}$. As can be deduced from Fig. \ref{fig:conv-weakly-f}, the convergence results for weakly convex objective functions are significantly different from the quardratic loss function. In the phase retrieval problem, the quantization bit exhibits much more influences than the convex case. In Fig. \ref{fig:comp}, we compare the proposed two-stage GDSRQ with early stoppping (ES-GDSRQ2) with standard SGD method and the naive GDSRQ (N-GDSRQ). Simulation results show that the proposed ES-GDSRQ2 has superior convergence speed over N-GDSRQ and outperforms standard SGD in terms of final accuracy. The pleasant performances of ES-GDSRQ2 motivates us that an appropriate choice of the stepsizes $\beta_t$ can remarkably boost the capability of GDSRQ. It still remains an open problem how to adaptively determine stepsizes in consideration of specific features of problems and algorithm evolution. 

\appendix
% #################### Proof #######################
\subsection{Proof of Lemma \ref{lma:convergence-of-production}}
\begin{proof}
  Firstly, the existence of $\lim_{k\rightarrow \infty} \prod_{t=0}^k (1-\delta x_t)$ directly follows the monotone convergence theorem. Next we proceed to prove the first half of the lemma by contradiction. Assume there exists some constant $0<p<1$ such that
  \begin{equation}
    \prod_{k=0}^\infty (1-\delta x_k) = p.
  \end{equation}
  Then we have
  \begin{equation}
    \sum_{k=0}^\infty \ln(1-\delta x_k) = \ln p,
  \end{equation}
  which indicates that 
  \begin{equation}
    \sum_{k=0}^\infty \ln\left(1 + \frac{\delta x_k}{1-\delta x_k}\right) = \ln \frac{1}{p}.
  \end{equation}
  Rewriting the left-hand side using Taylor's expansion with the Peano remainder $\theta(x_k) \in o(x_k^2)$, we obtain
  \begin{equation}
    \begin{aligned}
      &\sum_{k=0}^\infty \ln\left(1 + \frac{\delta x_k}{1-\delta x_k}\right) \\
      &= \sum_{k=0}^\infty \frac{\delta x_k}{1-\delta x_k} - \frac{1}{2} \left(\frac{\delta x_k}{1-\delta x_k}\right)^2 + \theta(x_k), \\
      &\ge \delta \sum_{k=0}^\infty x_k - \frac{\delta^2}{2} \sum_{k=0}^\infty \frac{x_k^2}{(1-\delta x_0)^2} + \sum_{k=0}^\infty \theta(x_k).
    \end{aligned}
  \end{equation}
  Since the series of $\{x_k\}$ is quadratically convergent, we have
  \begin{equation}
    \sum_{k=0}^\infty \theta(x_k) \le \frac{\delta^2}{2} \sum_{k=0}^\infty \frac{x_k^2}{(1-\delta x_0)^2} < \infty.
  \end{equation}
  Then it can be concluded that 
  \begin{equation}
    \sum_{k=0}^\infty x_k \le \ln \frac{1}{p} + \frac{\delta}{2} \sum_{k=0}^\infty \frac{x_k^2}{(1-\delta x_0)^2} - \frac{1}{\delta} \sum_{k=0}^\infty \theta(x_k) < \infty,
  \end{equation}
  which contradicts the divergence of $\sum_{k=0}^\infty x_k$. This completes the proof for the first half of the lemma. To derive the other half, we use $1-x\le e^{-x}$ for $0\le x\le 1$ as follows:
  \begin{equation}
    \sum_{k=0}^\infty \prod_{t=0}^k (1-\delta x_k) \le \sum_{k=0}^\infty \exp\left\{ -\delta\sum_{t=0}^k x_t \right\} .
  \end{equation}
  Since $\lim_{k\rightarrow \infty} x_k(k+1) = \infty$, we can choose $\varepsilon = 2/\delta$, a positive constant $F > 0$ and a corresponding $M=M(\varepsilon,F)$ such that $x_t(t+1)\ge \varepsilon$ and $\left|\sum_{i=0}^t \frac{1}{1+i} - \ln t \right| \le F$ for all $t\ge M$. Then we obtain
  \begin{equation}
    \label{eqn:beta-dealing}
    \begin{aligned}
      \sum_{k=M}^\infty \exp\left\{ -\delta\sum_{t=0}^k x_t \right\} &\le \sum_{k=M}^\infty \exp\left\{ -\delta\sum_{t=M}^k x_t \right\}, \\
      &= \sum_{k=M}^\infty \exp\left\{ -\delta\sum_{t=M}^k \frac{x_t(t+1)}{t+1}  \right\}, \\
      &\le \sum_{k=M}^\infty \exp\left\{ -\delta \varepsilon \sum_{t=M}^k \frac{1}{t+1} \right\}, \\
      &\le \sum_{k=M}^\infty e^{-2(\ln k - \ln M - 2F)}, \\
      &= M^2 e^{4F} \sum_{k=M}^\infty k^{-2} < \infty.
    \end{aligned}
  \end{equation}
  This indicates that 
  \begin{equation*}
    \sum_{k=0}^\infty \prod_{t=0}^k (1-\delta x_k) \le \sum_{k=0}^{M-1} e^{ -\delta\sum_{t=0}^k x_t } + \sum_{k=M}^\infty e^{ -\delta\sum_{t=0}^k x_t} < \infty,
  \end{equation*}
  which completes the proof.
\end{proof}

\subsection{Proof of Theorem \ref{thm:strongly-convex-convergence}}
\begin{proof}
  According to (\ref{eqn:gamma-ini}), we have 
  \begin{equation}
    \lim_{k\rightarrow \infty} \frac{\gamma_k^0}{\alpha_0} \le \lim_{k\rightarrow \infty} \sum_{t=0}^k \frac{\gamma_k^t \beta_t^2}{\alpha_t \beta_0^2 } = 0.
  \end{equation}
  From (\ref{eqn:gamma-constraint}), we can deduce that $\forall j\le k$
  \begin{equation}
    \frac{\gamma_k^j}{\alpha_j} = \sum_{t=1}^j \frac{\gamma_k^t}{\alpha_t} - \frac{\gamma_k^{t-1}}{\alpha_{t-1}} + \frac{\gamma_k^0}{\alpha_0},
  \end{equation}
  which indicates that
  \begin{equation}
    \lim_{k\rightarrow \infty} \frac{\gamma_k^j}{\alpha_j} \le \lim_{k\rightarrow \infty} \sum_{t=1}^j \left|\frac{\gamma_k^t}{\alpha_t} - \frac{\gamma_k^{t-1}}{\alpha_{t-1}} \right| + \frac{\gamma_k^0}{\alpha_0} \le 0.
  \end{equation}
  To find the relationship between $r(k+1)$ and $r(k)$, the first step is to estimate $- \frac{2\alpha_k}{N} \sum_{i=1}^N \mathbf{g}_i^T(\mathbf{x}_i(k)) (\mathbf{x}_i(k) - \mathbf{x}^*)$ by Lemma \ref{lma:x-avg-error}. Therefore, we have
  \begin{equation}
    \label{eqn:subgradient-term}
    \begin{aligned}
      &- \frac{2\alpha_k}{N} \sum_{i=1}^N \langle g_i(\mathbf{x}_i(k)), \mathbf{x}_i(k) - \mathbf{x}^* \rangle \\
      &\le - \frac{2\alpha_k}{N} \left[ \sum_{i=1}^N f_i(\mathbf{x}_i(k)) - f_i(\mathbf{x}^*) + \frac{\mu}{2} \|\mathbf{x}_i(k)-\mathbf{x}^*\|^2 \right], \\
      &= - \frac{2\alpha_k}{N} \left[ \sum_{i=1}^N f_i(\mathbf{x}_i(k)) - f_i(\bar{\mathbf{x}}(k)) + f_i(\bar{\mathbf{x}}(k)) - f_i(\mathbf{x}^*) \right] \\
      &\quad - \frac{\mu\alpha_k}{N} \sum_{i=1}^N \|\mathbf{x}_i(k)-\mathbf{x}^*\|^2.
    \end{aligned}
  \end{equation}
  Next we analyze each term respectively. Since $f_i$ is $L_i$-Lipchitz continuous, we have
  \begin{equation}
    |f_i(\mathbf{x}_i(k)) - f_i(\bar{\mathbf{x}}(k))| \le L_i \|\mathbf{x}_i(k) - \bar{\mathbf{x}}(k)\| = L_i \|\mathbf{y}_i(k)\|,
  \end{equation}
  and expanding this via a little trick, we obtain
  \begin{equation}
    \label{eqn:Cauchy-Schwatz-trick}
    2\alpha_k L_i \|\mathbf{y}_i(k)\| = 2\frac{\alpha_k L_i}{\sqrt{\beta_k}} \sqrt{\beta_k} \|\mathbf{y}_i(k)\| \le \frac{\alpha_k^2 L_i^2}{\beta_k}+\beta_k \|\mathbf{y}_i(k)\|^2.
  \end{equation}
  In addition, using the Cauchy-Schwartz inequality, we have
  \begin{equation}
    -\frac{1}{N}\sum_{i=1}^N \|\mathbf{x}_i(k)-\mathbf{x}^*\|^2 \le - \left\| \sum_{i=1}^N \frac{1}{N}\mathbf{x}_i(k)-\mathbf{x}^* \right\|^2 = -r_k.
  \end{equation}
  To derive $f_i(\bar{\mathbf{x}}(k)) - f_i(\mathbf{x}^*)$, we first fix some $\theta \in \mathcal{V}$:
  \begin{equation}
    f_i(\bar{\mathbf{x}}(k)) - f_i(\mathbf{x}^*) = f_i(\bar{\mathbf{x}}(k)) - f_i(\mathbf{x}_\theta(k))+ f_i(\mathbf{x}_\theta(k)) - f_i(\mathbf{x}^*).
  \end{equation}
  To estimate $f_i(\bar{\mathbf{x}}(k)) - f_i(\mathbf{x}_\theta(k))$, the Cauchy-Schwartz inequality is adopted similarly as (\ref{eqn:Cauchy-Schwatz-trick}):
  \begin{equation}
    -\frac{2\alpha_k}{N} \sum_{i=1}^N f_i(\bar{\mathbf{x}}(k)) - f_i(\mathbf{x}_\theta(k)) \le  \frac{L^2\alpha_k^2}{N\beta_k} + \beta_k \|\mathbf{Y}_k\|^2.
  \end{equation}
  Substituting the results above into (\ref{eqn:subgradient-term}), we obtain
  \begin{equation}
    \begin{aligned}
      &-\frac{2\alpha_k}{N} \sum_{i=1}^N \mathbf{g}_i^T(\mathbf{x}_i(k)) (\mathbf{x}_i(k) - \mathbf{x}^*) \le \frac{2L^2\alpha_k^2}{N\beta_k} \\
      &+ 2\beta_k \|\mathbf{Y}_k\|^2 - \mu\alpha_k r_k- 2\alpha_k \left[ f(\mathbf{x}_\theta(k)) - f(\mathbf{x}^*) \right].
    \end{aligned}
  \end{equation}
  The inequality in Lemma \ref{lma:x-avg-error} can be represented by
  \begin{equation}
    \begin{aligned}
      &\mathbb{E}[r_{k+1}|\mathscr{F}_k] \le \xi_k r_k + \frac{6L^2\alpha_k^2}{N(1-\beta_0)} + \frac{4L^2\alpha_k^2}{N\beta_k} \\
      &+ \Delta^2\beta_k^2 + \frac{4\beta_k \|\mathbf{Y}_k\|^2}{N} - 2\alpha_k[f(\mathbf{x}_\theta(k)) - f(\mathbf{x}^*)],
    \end{aligned}
  \end{equation}
  where $\xi_k = 1-\mu\alpha_k$. Note that without loss of generality, we assume for convenience of writing that
  \begin{equation}
    (1-\sigma_2)\beta_0 < 1, \quad \alpha_0\mu < 1.
  \end{equation}
  By taking expectation at both sides, the recursive upper bound formula can be written as follows:
  \begin{equation}
    \label{eqn:recursive-eta}
    \mathbb{E}[r_{t+1}] \le \xi_t \mathbb{E}[r_t] + h_{t} - 2\alpha_t \mathbb{E}[f(\mathbf{x}_\theta(t)) - f(\mathbf{x}^*)],
  \end{equation}
  where $h_{t}$ stands for the estimation drift error. Moving the last term to the left-hand side and multiplying $\gamma_k^t$, we obtain
  \begin{equation}
    \label{eqn:recursive-f}
    \begin{aligned}
      2\gamma_k^t \left( \mathbb{E}[f(\mathbf{x}_\theta(t))] - f(\mathbf{x}^*) \right)
      \le (\xi_t \mathbb{E}[r_t] - \mathbb{E}[r_{t+1}]+h_t)\frac{\gamma_k^t}{\alpha_t}, \\
      \le \mathbb{E}[r_t] \frac{\gamma_k^t \xi_t}{\alpha_t} - \mathbb{E}[r_{t+1}] \frac{\gamma_k^{t+1} \xi_{t+1}}{\alpha_{t+1}} + h_t \frac{\gamma_k^t}{\alpha_t},
    \end{aligned}
  \end{equation}
  where the last inequality results from $\xi_t < 1$ and (\ref{eqn:gamma-constraint}). Applying (\ref{eqn:recursive-f}) recursively for $t$ from $0$ to $k$, we obtain
  \begin{equation}
    \label{eqn:expansion-summing-f}
    \begin{aligned}
      &2 \left( \sum_{t=0}^k \gamma_k^t \mathbb{E}[f(\mathbf{x}_\theta(t))] - f(\mathbf{x}^*)  \right) \\
      &= \sum_{t=0}^k 2\gamma_k^t \left( \mathbb{E}[f(\mathbf{x}_\theta(t))] - f(\mathbf{x}^*) \right), \\
      &\le \mathbb{E}[r_0] \frac{\gamma_k^0\xi_0}{\alpha_0} - \mathbb{E}[r_{k+1}] \frac{\gamma_k^k}{\alpha_k} + \sum_{t=0}^k h_t \frac{\gamma_k^t}{\alpha_t}.
    \end{aligned}
  \end{equation}
  Next we use Lemma \ref{lma:x-deviation} to derive $\mathbb{E}\|\mathbf{Y}_k\|^2$. To be specific, (\ref{eqn:Yk-recursive}) can be written as a recursive formula:
  \begin{equation}
    \mathbb{E}\|\mathbf{Y}_{k+1}\|^2 \le \varphi_k \mathbb{E}\|\mathbf{Y}_k\|^2 + N\sigma_2^2 \Delta^2 \beta_k^2 + \frac{4\alpha_k^2 L^2(1+\beta_0)}{1-\varphi_k}, 
  \end{equation}
  where $\varphi_k = 1-(1-\sigma_2)\beta_k$. Via recursive expansion, we obtain
  \begin{equation}
    \begin{aligned}
      &\mathbb{E}\|\mathbf{Y}_{k+1}\|^2 \le \mathbb{E}\|\mathbf{Y}_0\|^2 \prod_{t=0}^k \varphi_t \\
      &+ \sum_{t=0}^k N\sigma_2^2 \Delta^2 \beta_t^2 \prod_{j=t+1}^k \varphi_j + \sum_{t=0}^k \frac{4\alpha_t^2 L^2(1+\beta_0)}{1-\varphi_t} \prod_{j=t+1}^k \varphi_j.
    \end{aligned}
  \end{equation}
  Then we continue to derive $\sum_{t=0}^k \mathbb{E}\|\mathbf{Y}_t\|^2$ as follows:
  \begin{equation}
    \begin{aligned}
      &\sum_{t=0}^k \frac{\gamma_k^t}{\alpha_t} \mathbb{E}\|\mathbf{Y}_t\|^2 \\
      &\le \mathbb{E}\|\mathbf{Y}_0\|^2 \sum_{t=0}^k \frac{\gamma_k^t}{\alpha_t} \prod_{i=0}^{t-1} \varphi_i + \sum_{t=0}^k \frac{\gamma_k^t}{\alpha_t} \sum_{i=0}^{t-1} N\sigma_2^2\Delta^2\beta_i^2 \prod_{j=i+1}^{t-1} \varphi_j \\
      &+ \sum_{t=0}^k \frac{\gamma_k^t}{\alpha_t} \sum_{i=0}^{t-1} \frac{4\alpha_i^2 L^2(1+\beta_0)}{1-\varphi_i} \prod_{j=i+1}^{t-1} \varphi_j.
    \end{aligned}
  \end{equation}
  According to Lemma \ref{lma:x-deviation}, we expand the term concerning $\sum_{t=0}^k \mathbb{E}\|\mathbf{Y}_t\|^2$ as (\ref{eqn:coupling-term}) at the top of the next page, where $\omega_k^t = \gamma_k^{t+1}/\alpha_{t+1} - \gamma_k^t/\alpha_t$ for $0<t<k$, $\omega_k^0 = \gamma_k^0/\alpha_0$ and $\omega_k^k = \gamma_k^k/\alpha_k$. 
  \begin{figure*}[!t]
    \normalsize
    \begin{equation}
      \label{eqn:coupling-term}
      \begin{aligned}
      \sum_{t=0}^k \frac{4\beta_t\gamma_k^t}{N\alpha_t} \mathbb{E}\|\mathbf{Y}_t\|^2 &\le \sum_{t=0}^k \frac{4\beta_t\gamma_k^t}{N\alpha_t} \frac{\mathbb{E}\|\mathbf{Y}_t\|^2 - \mathbb{E}\|\mathbf{Y}_{t+1}\|^2}{1-\varphi_t} + \sum_{t=0}^k \frac{4\beta_t\gamma_k^t}{N\alpha_t} \left[\frac{N\sigma_2^2\Delta^2\beta_t^2}{1-\varphi_t} + \frac{4\alpha_t^2L^2(\beta_0+1)}{(1-\varphi_t)^2} \right], \\
      &= \frac{4}{N(1-\sigma_2)}\sum_{t=0}^{k+1} \omega_k^t \mathbb{E}\|\mathbf{Y}_t\|^2 + \sum_{t=0}^k \frac{\gamma_k^t}{\alpha_t} \left[\frac{4 \sigma_2^2\Delta^2\beta_t^2}{1-\sigma_2} + \frac{16\alpha_t^2L^2(\beta_0+1)}{N(1-\sigma_2)^2\beta_t} \right], \\
      &\le \frac{4}{N(1-\sigma_2)} \mathbb{E}\|\mathbf{Y}_0\|^2 \sum_{t=0}^k |\omega_k^t| \prod_{i=0}^{t-1} \varphi_i + \sum_{t=0}^k |\omega_k^t| \sum_{i=0}^{t-1} \left[ \frac{4 \sigma_2^2\Delta^2\beta_i^2}{1-\sigma_2} + \frac{16\alpha_i^2L^2(\beta_0+1)}{N(1-\sigma_2)^2\beta_i} \right] \prod_{j=i+1}^{t-1} \varphi_j  \\
      &+ \sum_{t=0}^k \frac{\gamma_k^t}{\alpha_t} \left[\frac{4 \sigma_2^2\Delta^2\beta_t^2}{1-\sigma_2} + \frac{16\alpha_t^2L^2(\beta_0+1)}{N(1-\sigma_2)^2\beta_t} \right],
      \end{aligned}
    \end{equation}
    \hrulefill
    \vspace*{2pt}
  \end{figure*}
  To estimate (\ref{eqn:coupling-term}), we derive each term separately. First we have
  \begin{equation}
    \sum_{t=1}^k \omega_k^t \prod_{i=0}^{t-1} \varphi_i \le \sum_{t=1}^k \omega_k^t \exp\left\{ -\delta\sum_{i=0}^{t-1}\beta_i \right\},
  \end{equation}
  where $\delta = 1-\sigma_2$. Then we \textbf{adopt similar derivation in the proof of Lemma \ref{lma:convergence-of-production}} as (\ref{eqn:beta-dealing}). Recall that we have assumed $\lim_{k\rightarrow \infty} (k+1) \beta_k =\infty$ and $\lim_{k\rightarrow \infty}k^{-1}/\alpha_k\le C$. We can choose $\varepsilon_1 = 2/\delta$, $\varepsilon_2>0$, $F>0$, and a corresponding integer $M_1=M_1(\varepsilon_1,\varepsilon_2,F)$ such that $t\beta_t \ge \varepsilon_1$, $\left|\sum_{i=0}^{t-1} \frac{1}{i+1} - \ln t \right| \le F$, and $(t-1)^{-2} \le \varepsilon_2$ for all $t\ge M_1$. Then we obtain the following upper bound for $k>M_1+1$:
  \begin{equation*}
    \begin{aligned}
      \sum_{t=M_1+1}^k \omega_k^t e^{-\delta \sum_{i=0}^{t-1} \beta_i} &\le M_1^2 e^{4F} \sum_{t=M_1+1}^k \omega_k^t (t-1)^{-2}, \\
      &\le \varepsilon_2 M_1^2 e^{4F} \sum_{t=M_1+1}^k \omega_k^t, 
    \end{aligned}
  \end{equation*}
  which by (\ref{eqn:assumption-stepsizes1}) indicates that 
  \begin{equation}
    \label{eqn:convergence-1-term}
    \begin{aligned}
      &\lim_{k\rightarrow \infty} \sum_{t=0}^k \omega_k^t \prod_{i=0}^{t-1} \varphi_i \le \lim_{k\rightarrow \infty} \sum_{t=1}^k \omega_k^t \exp\left\{ -\delta\sum_{i=0}^{t-1}\beta_i \right\}\\
      &\le \lim_{k\rightarrow \infty} \omega_k^0 + \sum_{t=1}^{M_1} \omega_k^t + \varepsilon_2 M_1^2e^{4F} \sum_{t=M_1+1}^k \omega_k^t, \\
      &\le 0 + 0 + 0 = 0.
    \end{aligned}
  \end{equation}
  Next we continue to derive the second term:
  \begin{equation}
    \sum_{t=0}^k \omega_k^t \sum_{i=0}^{t-1} \beta_i^2 \prod_{j=i+1}^{t-1} \varphi_j \le \sum_{t=0}^k \omega_k^t \sum_{i=0}^{t-1} \beta_i^2 e^{ -\delta \sum_{j=i+1}^{t-1} \beta_j }.
  \end{equation}
  Recall the assumptions in (\ref{eqn:assumption-stepsizes1}). We choose $\varepsilon_3$, $\varepsilon_4$, and a corresponding integer $M > M_1$ such that for all $k\ge M$,
  \begin{equation}
    \label{eqn:convergence-separate-terms}
    \sum_{t=M+2}^k \frac{(t+1)^2\beta_t^2}{k^2} \le \varepsilon_3, \ \sum_{t=M+2}^k \frac{ t^2 \alpha_t^2 }{k^2 \beta_t} \le \varepsilon_4.
  \end{equation}
  For $t>M+1$, we have the following relationship which is different from (\ref{eqn:beta-dealing}):
  \begin{equation}
    \sum_{i=M}^{t-1} \beta_i^2 e^{ -\delta \sum_{j=i+1}^{t-1} \beta_j } \le e^{4F} \sum_{i=M}^{t-1} \beta_i^2 \left(\frac{i-1}{t+1}\right)^2, 
  \end{equation}
  which can be further written as
  \begin{equation}
    \begin{aligned}
      \omega_k^t \sum_{i=M}^{t-1} \beta_i^2 \prod_{j=i+1}^{t-1} \varphi_j &\le e^{4F} \omega_k^t \sum_{i=M}^{t-1} \frac{(i+1)^2\beta_i^2}{t^2}, \\
      &\le e^{4F}\varepsilon_3 \omega_k^t.
    \end{aligned}
  \end{equation}
  Therefore, we obtain
  \begin{equation}
    \lim_{k\rightarrow \infty} \sum_{t=M+2}^k \omega_k^t \sum_{i=M}^{t-1} \beta_i^2 \prod_{j=i+1}^{t-1} \varphi_j = 0.
  \end{equation}
  Moreover, we have
  \begin{equation}
    \begin{aligned}
      \omega_k^t \sum_{i=0}^{M-1} \beta_i^2 \prod_{j=i+1}^{t-1} \varphi_j &\le \omega_k^t \sum_{i=0}^{M-1} \beta_i^2 \prod_{j=M}^{t-1} \varphi_j, \\
      &\le M^3 e^{4F} \omega_k^t (t-1)^{-2}.
    \end{aligned}
  \end{equation}
  By taking limit of the sum at both sides, we obtain
  \begin{equation}
    \lim_{k\rightarrow \infty} \sum_{t=M+2}^k \omega_k^t \sum_{i=0}^{M-1} \beta_i^2 \prod_{j=i+1}^{t-1} \varphi_j = 0.
  \end{equation}
  In terms of the finite sum part, we have
  \begin{equation}
    \lim_{k\rightarrow \infty} \omega_k^t = 0 \Rightarrow \lim_{k\rightarrow \infty} \sum_{t=0}^{M+1} \omega_k^t \sum_{i=0}^{t-1} \beta_i^2 \prod_{j=i+1}^{t-1} \varphi_j = 0.
  \end{equation}
  Therefore, we finally obatin
  \begin{equation}
    \label{eqn:convergence-2-1-term}
    \lim_{k\rightarrow \infty} \sum_{t=0}^{k} \omega_k^t \sum_{i=0}^{t-1} \beta_i^2 \prod_{j=i+1}^{t-1} \varphi_j = 0.
  \end{equation}
  Similar derivation can be made once again to prove
  \begin{equation}
    \label{eqn:convergence-2-2-term}
    \lim_{k\rightarrow \infty} \sum_{t=0}^{k} \omega_k^t \sum_{i=0}^{t-1} \frac{\alpha_i^2}{\beta_i} \prod_{j=i+1}^{t-1} \varphi_j = 0,
  \end{equation}
  which is omitted here due to page limit. The limit of the last term in (\ref{eqn:coupling-term}) is direct to obtain as follows:
  \begin{subequations}
    \begin{equation}
      \label{eqn:convergence-3-1-term}
      \lim_{k\rightarrow \infty} \sum_{t=0}^k \frac{\gamma_k^t}{\alpha_t}\frac{4\sigma_2^2\Delta^2\beta_t^2}{1-\sigma_2} = \frac{4\sigma_2^2\Delta^2}{1-\sigma_2} \lim_{k\rightarrow \infty}\sum_{t=0}^k \frac{\gamma_k^t}{\alpha_t} \beta_t^2 =0,
    \end{equation}
    \begin{equation}
      \label{eqn:convergence-3-2-term}
      \frac{16L^2(\beta_0+1)}{N(1-\sigma_2)^2} \lim_{k\rightarrow \infty} \sum_{t=0}^k \frac{\gamma_k^t \alpha_t}{\beta_t}  = 0.
    \end{equation}
  \end{subequations}
  Combining (\ref{eqn:convergence-1-term}), (\ref{eqn:convergence-2-1-term}), (\ref{eqn:convergence-2-2-term}), (\ref{eqn:convergence-3-1-term}), and (\ref{eqn:convergence-3-2-term}), we obtain
  \begin{equation}
    \lim_{k\rightarrow \infty} \sum_{t=0}^k \frac{4\beta_t\gamma_k^t}{N\alpha_t} \mathbb{E}\|\mathbf{Y}_t\|^2 = 0.
  \end{equation}
  Applying (\ref{eqn:convergence-3-1-term}) and (\ref{eqn:convergence-3-2-term}) again, we have
  \begin{equation}
    \lim_{k\rightarrow \infty}\sum_{t=0}^k h_t\frac{\gamma_k^t}{\alpha_t} = 0.
  \end{equation}
  Due to the strong convexity of $f$ and using Jensen's inequality, we obtain for all $k$ and any $\theta \in \mathcal{V}$
  \begin{equation}
    \begin{aligned}
      \frac{\mu}{2}\mathbb{E}\|\mathbf{z}_\theta(k) - \mathbf{x}^*\|^2 &\le \mathbb{E}[f(\mathbf{z}_i(k))] - f(\mathbf{x}^*), \\
      &\le \sum_{t=0}^k \gamma_k^t \mathbb{E}[f(\mathbf{x}_\theta(k))] - f(\mathbf{x}^*).
    \end{aligned}
  \end{equation}
  Based on this, we obtain
  \begin{equation}
    \lim_{k\rightarrow \infty} \mathbb{E}\|\mathbf{z}_\theta(k) - \mathbf{x}^*\|^2 = 0,
  \end{equation}
  which completes the proof.
\end{proof}

\subsection{Proof of Proposition \ref{prop:convex-convergence}}
\begin{proof}
	The last term in Lemma \ref{lma:x-avg-error} can be written as
	\begin{equation*}
	  \begin{aligned}
      &- \frac{2\alpha_k}{N} \sum_{i=1}^N \langle \mathbf{g}_i(\mathbf{x}_i(k)), \mathbf{x}_i(k) - \mathbf{x}^* \rangle \\
      &\le - \frac{2\alpha_k}{N} \left[ \sum_{i=1}^N f_i(\mathbf{x}_i(k)) - f_i(\mathbf{x}^*)\right], \\
      &= - \frac{2\alpha_k}{N} \left[ \sum_{i=1}^N f_i(\mathbf{x}_i(k)) - f_i(\bar{\mathbf{x}}(k)) + f_i(\bar{\mathbf{x}}(k)) - f_i(\mathbf{x}^*) \right].
    \end{aligned}
	\end{equation*}
	By adopting Cauchy-Schwartz ineuqality and based on the fact that $f_i$ is $L$-Lipchitz continuous, we obtain
	\begin{subequations}
		\begin{equation*}
			|f_i(\mathbf{x}_i(k)) - f_i(\bar{\mathbf{x}}(k))| \le L_i \|\mathbf{x}_i(k) - \bar{\mathbf{x}}(k)\| = L_i \|\mathbf{y}_i(k)\|,
		\end{equation*}
		\begin{equation*}
			2\alpha_kL_i \|\mathbf{y}_i(k)\| = 2\frac{\alpha_k L_i}{\sqrt{\beta_k}} \sqrt{\beta_k} \|\mathbf{y}_i(k)\| \le \frac{\alpha_k^2 L_i^2}{\beta_k}+\beta_k \|\mathbf{y}_i(k)\|^2.
		\end{equation*}
		\begin{equation*}
			f_i(\bar{\mathbf{x}}(k)) - f_i(\mathbf{x}^*) = f_i(\bar{\mathbf{x}}(k)) - f_i(\mathbf{x}_\theta(k))+ f_i(\mathbf{x}_\theta(k)) - f_i(\mathbf{x}^*).
		\end{equation*}
	\end{subequations}
	Then by Lemma \ref{lma:x-avg-error}, we have
	\begin{equation}
	  \begin{aligned}
      &\mathbb{E}[r_{k+1}|\mathscr{F}_k] \le r_k + \frac{6L^2\alpha_k^2}{N(1-\beta_0)} + \frac{4L^2\alpha_k^2}{N\beta_k} + \Delta^2\beta_k^2 \\
      & + \frac{4\beta_k \|\mathbf{Y}_k\|^2}{N} - 2\alpha_k[f(\mathbf{x}_\theta(k)) - f(\mathbf{x}^*)].
    \end{aligned}
    \label{eqn:rk-recursive}
	\end{equation}
	By taking expectation at both sides and rearranging the ineuqality, we obtain
	\begin{equation*}
		\begin{aligned}
			2\gamma_k^t \left( \mathbb{E}[f(\mathbf{x}_\theta(t))] - f(\mathbf{x}^*) \right)  \le (\mathbb{E}[r_t] - \mathbb{E}[r_{t+1}]+h_{t})\frac{\gamma_k^t}{\alpha_t}, \\
			\le \mathbb{E}[r_t]\frac{\gamma_k^t}{\alpha_t} - \mathbb{E}[r_{t+1}]\frac{\gamma_k^{t+1}}{\alpha_{t+1}} +h_{t} \frac{\gamma_k^t}{\alpha_t},
		\end{aligned}
	\end{equation*}
	where the second inequality results from $\gamma_k^t/\alpha_t \ge \gamma_k^{t+1}/\alpha_{t+1}$. Summing both sides recursively, we have
	\begin{equation}
		\begin{aligned}
      &2 \left( \sum_{t=0}^k \gamma_k^t \mathbb{E}[f(\mathbf{x}_\theta(t))] - f(\mathbf{x}^*)  \right) \\
      &= \sum_{t=0}^k 2\gamma_k^t \left( \mathbb{E}[f(\mathbf{x}_\theta(t))] - f(\mathbf{x}^*) \right), \\
      &\le \mathbb{E}[r_0] \frac{\gamma_k^0}{\alpha_0} - \mathbb{E}[r_{k+1}] \frac{\gamma_k^k}{\alpha_k} + \sum_{t=0}^k h_t \frac{\gamma_k^t}{\alpha_t},
    \end{aligned}
	\end{equation}
	which is in consistency with (\ref{eqn:expansion-summing-f}). Based on similar derivation of the proof of Theorem 1, we have
	\begin{equation}
		\lim_{k\rightarrow \infty} \mathbb{E}\|\mathbf{z}_\theta(k) - \mathbf{x}^*\|^2 = 0,
	\end{equation}
	which completes the proof.
\end{proof}

\subsection{Proof of Proposition \ref{prop:convergence-rates-special-case}}
\begin{proof}
  Denote by $\Phi_\gamma$ the approximation to the denominator of $\gamma_k^t$ for sufficiently large $k$, which is expressed as
  \begin{equation}
    \Phi_\gamma(k) = \left\{
    \begin{aligned}
      \ln(k+1), && \lambda_\gamma = 1, \\
      (k+1)^{1-\lambda_\gamma}, && \lambda_\gamma < 1.
    \end{aligned}
    \right.
  \end{equation}
  Such approximation is non-expansive since we have $\forall k > 1$
  \begin{subequations}
    \begin{equation}
      \sum_{t=0}^k (t+1)^{-\lambda_\gamma} \le \int_{0}^k (t+1)^{-\lambda_\gamma} dt \le 2 \Phi_\gamma(k),
    \end{equation}
    \begin{equation}
      \sum_{t=0}^k (t+1)^{-\lambda_\gamma} \ge \frac{1}{4} \int_{0}^k (t+1)^{-\lambda_\gamma} dt \ge \frac{1}{8} \Phi_\gamma(k).
    \end{equation}
  \end{subequations}
  Based on (\ref{eqn:rk-recursive}), we have
  \begin{equation}
    \begin{aligned}
      &\mathbb{E}[r_{t+1}] \le \mathbb{E}[r_t] + \frac{6L^2\alpha_0^2}{N(1-\beta_0)}\frac{1}{(t+1)^{2\lambda_\alpha}} \\
      &+ \frac{4L^2\alpha_0^2}{N\beta_0}\frac{1}{(t+1)^{2\lambda_\alpha - \lambda_\beta}} + \frac{\Delta^2\beta_0^2}{(t+1)^{2\lambda_\beta}} \\
      &+ \frac{4\beta_0 \mathbb{E}\|\mathbf{Y}_t\|^2}{N(t+1)^{\lambda_\beta}} - \frac{2\alpha_0}{(t+1)^{\lambda_\alpha}}[f(\mathbf{x}_\theta(k)) - f(\mathbf{x}^*)].
    \end{aligned}
  \end{equation}
  Moving the last term to LHS and applying the fact $\lambda_\alpha \ge \lambda_\gamma$, we obtain the following approximate result for large $k$:
  \begin{equation}
    \begin{aligned}
      & \ 2 \left( \sum_{t=0}^k \gamma_k^t \mathbb{E}[f(\mathbf{x}_\theta(t))] - f(\mathbf{x}^*)  \right) \\
      &= \sum_{t=0}^k 2\gamma_k^t \left( \mathbb{E}[f(\mathbf{x}_\theta(t))] - f(\mathbf{x}^*) \right), \\
      &\le  \frac{\mathbb{E}[r_0]}{\alpha_0 \Phi_\gamma(k)} + \frac{\mathbb{E}[r_{k+1}] (k+1)^{\lambda_\alpha-\lambda_\gamma}}{\alpha_0 \Phi_\gamma(k)} + \sum_{t=0}^k h_t \frac{\gamma_k^t}{\alpha_t},
    \end{aligned}
  \end{equation}
  By the definition of $\Phi_\gamma$, we have
  \begin{equation}
    \label{eqn:phi-gamma}
    \frac{(k+1)^{\lambda_\alpha-\lambda_\gamma}}{\Phi_\gamma(k)} = \left\{
    \begin{aligned}
      \frac{(k+1)^{\lambda_\alpha-1}}{\ln(k+1)}, && \lambda_\gamma=1,\\
      (k+1)^{\lambda_\alpha-1}, && \lambda_\gamma<1.
    \end{aligned}
    \right.
  \end{equation}
  Applying (\ref{eqn:phi-gamma}) and the limit of $\mathbb{E}[r_k]$ in (\ref{eqn:rk-limit}), we obtain
  \begin{equation}
    \lim_{k\rightarrow \infty} \frac{\mathbb{E}[r_0]}{\alpha_0 \Phi_\gamma(k)} = 0, \ \lim_{k\rightarrow \infty} \frac{\mathbb{E}[r_{k+1}] (k+1)^{\lambda_\alpha-\lambda_\gamma}}{\alpha_0 \Phi_\gamma(k)} = 0.
  \end{equation}
  We proceed to derive the items in the last term separately. For all $k$, we have
  \begin{subequations}
    \begin{equation*}
      \sum_{t=0}^k \frac{6L^2\alpha_t\gamma_k^t}{N(1-\beta_0)} \le \int_0^k \frac{6L^2\alpha_t\gamma_k^t}{N(1-\beta_0)} dt \le \frac{96L^2}{N(1-\beta_0)} \frac{\Phi_{\alpha+\gamma}(k)}{\Phi_\gamma(k)},
    \end{equation*}
    \begin{equation*}
      \sum_{t=0}^k \frac{4L^2\alpha_t\gamma_k^t}{N\beta_t} \le \int_0^k \frac{4L^2\alpha_t\gamma_k^t}{N\beta_t} dt \le \frac{64L^2\alpha_0}{N\beta_0} \frac{\Phi_{\alpha+\gamma-\beta}(k)}{\Phi_\gamma(k)},
    \end{equation*}
    \begin{equation*}
      \sum_{t=0}^k \frac{\Delta^2\beta_t^2\gamma_k^t}{\alpha_t} \le \int_0^k \frac{\Delta^2\beta_t^2\gamma_k^t}{\alpha_t} dt \le \frac{16\Delta^2\beta_0^2}{\alpha_0} \frac{\Phi_{2\beta+\gamma-\alpha}(k)}{\Phi_\gamma(k)}.
    \end{equation*}
  \end{subequations}
  Next we derive the term containing $\mathbb{E}\|\mathbf{Y}_t\|^2$. Firstly, we notice that $1-(1-\sigma_2)\beta_k$ can be written as
  \begin{equation}
    1-(1-\sigma_2)\beta_k = 1- \frac{(1-\sigma_2)\beta_0}{(k+1)^{\lambda_\beta}} \le \left(\frac{k}{k+1}\right)^{\lambda_\beta},
  \end{equation}
  where ``$\le$'' results from $1-(1-\sigma_2)\beta_0 < 0$. According to Lemma \ref{lma:x-deviation}, $\mathbb{E}\|\mathbf{Y}_{t+1}\|^2$ can be expanded as
  \begin{equation*}
    \begin{aligned}
      & \ \mathbb{E}\|\mathbf{Y}_{k+1}\|^2 \\ 
      &\le \frac{k^{\lambda_\beta}}{(k+1)^{\lambda_\beta}} \mathbb{E}\|\mathbf{Y}_k \|^2 + N\sigma_2^2 \Delta^2 \beta_k^2 + \frac{4\alpha_k^2 L^2 (\beta_0+1)}{\beta_k (1-\sigma_2)}, \\
      &\le \frac{1}{(k+1)^{\lambda_\beta}} \left[N\sigma_2^2 \Delta^2 \sum_{t=0}^k \beta_t^2  + \frac{4 L^2 (\beta_0+1)}{(1-\sigma_2)} \sum_{t=0}^k \frac{\alpha_t^2}{\beta_t} \right], \\
      &\le \frac{N\sigma_2^2 \Delta^2\beta_0^2}{(k+1)^{3\lambda_\beta - 1}} + \frac{4 \alpha_0^2 L^2 (\beta_0+1)}{\beta_0(1-\sigma_2)} \frac{1}{(k+1)^{2\lambda_\alpha-1}}.
    \end{aligned}
  \end{equation*}
  Then we have
  \begin{equation}
    \label{eqn:Yt-recur}
    \begin{aligned}
      \sum_{t=0}^k \frac{4\beta_0 \mathbb{E}\|\mathbf{Y}_t\|^2}{N(t+1)^{\lambda_\beta}} \frac{\gamma_k^t}{\alpha_t} \le \sum_{t=0}^k \frac{4\beta_0\sigma_2^2\Delta^2}{\alpha_0 (t+1)^{4\lambda_\beta +\lambda_\gamma-\lambda_\alpha-1}\Phi_\gamma} \\ 
      + \sum_{t=0}^k \frac{16L^2\alpha_0(\beta_0+1)}{N(1-\sigma_2)(t+1)^{\lambda_\alpha+\lambda_\beta+\lambda_\gamma-1}\Phi_\gamma}.
    \end{aligned}
  \end{equation}

  \textbf{Case \bigroman{1}}: $\lambda_\gamma=1$. Then we have
  \begin{subequations}
    \begin{equation}
      \sum_{t=0}^k \frac{6L^2\alpha_t\gamma_k^t}{N(1-\beta_0)} \le \frac{96L^2}{N(1-\beta_0)} \frac{(k+1)^{-\lambda_\alpha}}{\ln(k+1)},
    \end{equation}
    \begin{equation}
      \sum_{t=0}^k \frac{4L^2\alpha_t\gamma_k^t}{N\beta_t} \le \frac{64L^2\alpha_0}{N\beta_0} \frac{(k+1)^{\lambda_\beta-\lambda_\alpha}}{\ln(k+1)},
    \end{equation}
    \begin{equation}
      \sum_{t=0}^k \frac{\Delta^2\beta_t^2\gamma_k^t}{\alpha_t} \le \frac{16\Delta^2\beta_0^2}{\alpha_0} \frac{(k+1)^{\lambda_\alpha-2\lambda_\beta}}{\ln(k+1)},
    \end{equation}
  \end{subequations}
  and (\ref{eqn:Yt-recur}) can be written as
  \begin{equation}
    \begin{aligned}
      \sum_{t=0}^k \frac{4\beta_0 \mathbb{E}\|\mathbf{Y}_t\|^2}{N(t+1)^{\lambda_\beta}} \frac{\gamma_k^t}{\alpha_t} \le \frac{4\beta_0\sigma_2^2\Delta^2}{\alpha_0} \frac{(k+1)^{1+\lambda_\alpha-4\lambda_\beta}}{\ln(k+1)} \\ 
      + \frac{16L^2\alpha_0(\beta_0+1)}{N(1-\sigma_2)} \frac{(k+1)^{1-\lambda_\alpha-\lambda_\beta}}{\ln(k+1)}.
    \end{aligned}
  \end{equation}

  \textbf{Case \bigroman{2}}: $\lambda_\gamma<1$. Then we have
  \begin{subequations}
    \begin{equation}
      \sum_{t=0}^k \frac{6L^2\alpha_t\gamma_k^t}{N(1-\beta_0)} \le \frac{96L^2}{N(1-\beta_0)} (k+1)^{-\lambda_\alpha},
    \end{equation}
    \begin{equation}
      \sum_{t=0}^k \frac{4L^2\alpha_t\gamma_k^t}{N\beta_t} \le \frac{64L^2\alpha_0}{N\beta_0} (k+1)^{\lambda_\beta-\lambda_\alpha},
    \end{equation}
    \begin{equation}
      \sum_{t=0}^k \frac{\Delta^2\beta_t^2\gamma_k^t}{\alpha_t} \le \frac{16\Delta^2\beta_0^2}{\alpha_0} (k+1)^{\lambda_\alpha-2\lambda_\beta},
    \end{equation}
  \end{subequations}
  and (\ref{eqn:Yt-recur}) can be written as
  \begin{equation}
    \begin{aligned}
      \sum_{t=0}^k \frac{4\beta_0 \mathbb{E}\|\mathbf{Y}_t\|^2}{N(t+1)^{\lambda_\beta}} \frac{\gamma_k^t}{\alpha_t} \le \frac{4\beta_0\sigma_2^2\Delta^2}{\alpha_0} (k+1)^{1+\lambda_\alpha-4\lambda_\beta} \\ 
      + \frac{16L^2\alpha_0(\beta_0+1)}{N(1-\sigma_2)} (k+1)^{1-\lambda_\alpha-\lambda_\beta}.
    \end{aligned}
  \end{equation}

  Combining the two cases above, the convergence rate $\chi$ determined by the minimum loss decay rate is expressed as
  \begin{equation}
    \chi = \max\{\lambda_\beta-\lambda_\alpha, 1+\lambda_\alpha-4\lambda_\beta\}.
  \end{equation}
  To minimize $\chi$, it is sufficient to set
  \begin{equation}
    \lambda_\alpha=\frac{5}{2}\lambda_\beta - \frac{1}{2},
  \end{equation}
  and we obtain
  \begin{equation}
    \min \chi = \frac{1}{2}-\frac{3}{2}\lambda_\beta,
  \end{equation}
  which completes the proof.
\end{proof}

\subsection{Proof of Lemma \ref{lma:weakly-convex-Lipchitz}}
\begin{proof}
  Denoting $\mathbf{u} = \text{prox}_{\theta \varphi} (\mathbf{x})$ and $\mathbf{v} = \text{prox}_{\theta \varphi} (\mathbf{y})$, by the definition of proximal mapping, it follows that
  \begin{equation}
    \frac{\mathbf{x} - \mathbf{u}}{\theta} \in \partial \varphi(\mathbf{u}), \ \frac{\mathbf{y}-\mathbf{v}}{\theta} \in \partial \varphi(\mathbf{v}).
  \end{equation}
  Thus, by the subgradient inequality,
  \begin{subequations}
    \begin{equation}
      \varphi(\mathbf{v}) \ge \varphi(\mathbf{u}) + \frac{1}{\theta} \langle \mathbf{x}-\mathbf{u}, \mathbf{v} - \mathbf{u} \rangle - \frac{\rho}{2} \|\mathbf{v} - \mathbf{u}\|^2,
    \end{equation}
    \begin{equation}
      \varphi(\mathbf{u}) \ge \varphi(\mathbf{v}) + \frac{1}{\theta} \langle \mathbf{y}-\mathbf{v}, \mathbf{u} - \mathbf{v} \rangle - \frac{\rho}{2} \|\mathbf{v} - \mathbf{u}\|^2.
    \end{equation}
  \end{subequations}
  Summing the two inequalities, we have
  \begin{equation}
    (1-\rho \theta) \|\mathbf{v} - \mathbf{u}\|^2 \le \langle \mathbf{x}-\mathbf{y}, \mathbf{u}-\mathbf{v} \rangle.
  \end{equation}
  Using the Cauchy-Swarchtz ineuqality, it follows that
  \begin{equation}
    \|\mathbf{u} - \mathbf{v}\| \le \frac{1}{1- \theta\rho} \|\mathbf{x}-\mathbf{y}\|,
  \end{equation}
  which completes the proof.
\end{proof}

\subsection{Proof of Theorem \ref{thm:weakly-convex-convergence}}
\begin{proof}
  The proof follows a similar methodology to analysis of convex functions. The main distinction between the methods applied in the proof and typical methods used in convex analysis lies in the analysis of the Moreau envelope. To discuss the convergence property of $\varphi_{i,\theta}(\mathbf{x}_{i,t})$, it is significant to exploit the relationships between $\varphi_{i,\theta}(\mathbf{x}_{i,t})$, $\varphi_{i,\theta}(\mathbf{x}_{i,t+1})$ and $\varphi_{i,\theta}(\mathbf{v}_{i,t})$. Without special statement, the term ``a.s.'' (almost sure) is omitted in the context for brevity of writing. Letting $\varphi_{\theta}(\mathbf{x}) = \frac{1}{N}\sum_{i=1}^N \varphi_{i,\theta}(\mathbf{x})$ and recalling $\varphi_{i,\theta}: \mathbf{x} \mapsto f_i(\hat{\mathbf{x}}) + \frac{1}{2\theta} \|\mathbf{x} - \hat{\mathbf{x}}\|^2$, we expand $\varphi_{\theta}(\mathbf{x}_{i,t+1})$ as
  \begin{equation}
    \label{eqn:varphi-xt1-expansion}
    \begin{aligned}
      & \ \varphi_{\theta}(\mathbf{x}_{i,t+1}) \\
      &= \min_{\mathbf{y}} \left\{ f(\mathbf{y}) + \frac{1}{2\theta} \|\mathbf{y} - \mathbf{x}_{i,t+1}\|^2 \right\},\\
      &\le f(\hat{\mathbf{v}}_{i,t}) + \frac{1}{2\theta} \left\| \hat{\mathbf{v}}_{i,t} - \mathcal{P}_{\mathcal{X}} [\mathbf{v}_{i,t} - \alpha_t \mathbf{g}_{i,t}] \right\|^2, \\
      &= f(\hat{\mathbf{v}}_{i,t}) + \frac{1}{2\theta} \left\| \mathcal{P}_{\mathcal{X}} [\mathbf{v}_{i,t} - \hat{\mathbf{v}}_{i,t} - \alpha_t \mathbf{g}_{i,t}] \right\|^2, \\
      &\le f(\hat{\mathbf{v}}_{i,t}) + \frac{1}{2\theta} \left\| \mathbf{v}_{i,t} - \hat{\mathbf{v}}_{i,t} - \alpha_t \mathbf{g}_{i,t} \right\|^2,\\ 
      &\le \varphi_{\theta}(\mathbf{v}_{i,t}) + \frac{\alpha_t^2}{2\theta} \|\mathbf{g}_{i,t}\|^2 + \frac{\alpha_t}{\theta}\langle \hat{\mathbf{v}}_{i,t} - \mathbf{v}_{i,t}, \mathbf{u}_{i,t} \rangle \\  
      & \quad + \frac{\alpha_t}{\theta}\langle \hat{\mathbf{v}}_{i,t} - \mathbf{v}_{i,t}, \mathbf{g}_{i,t} - \mathbf{u}_{i,t} \rangle,\\
      &\le \varphi_{\theta}(\mathbf{v}_{i,t}) + \frac{\alpha_t^2L^2}{2\theta} + \frac{\alpha_t}{\theta} \left[ f_i(\hat{\mathbf{v}}_{i,t}) - f_i(\mathbf{v}_{i,t}) \right] \\
      & \quad + \frac{\rho \alpha_t}{2\theta} \| \hat{\mathbf{v}}_{i,t} - \mathbf{v}_{i,t} \|^2 + \frac{\alpha_t}{\theta}\langle \hat{\mathbf{v}}_{i,t} - \mathbf{v}_{i,t}, \mathbf{g}_{i,t} - \mathbf{u}_{i,t} \rangle,
    \end{aligned}
  \end{equation}
  where $\mathbf{u}_{i,t}$ denotes the subgradient of $f_i$ evaluated at $\mathbf{v}_{i,t}$, and the last inequality stems from that $f_i$ is $\rho$-weakly convex and $L$-Lipchitz. Based on Assumption \ref{asp:lower-bounded}, we have
  \begin{equation}
    \begin{aligned}
      & \frac{\alpha_t}{\theta} \langle \hat{\mathbf{v}}_{i,t} - \mathbf{v}_{i,t}, \mathbf{g}_{i,t} - \mathbf{u}_{i,t} \rangle \\
      &\le \nu \frac{\alpha_t}{\theta} \|\hat{\mathbf{v}}_{i,t} - \mathbf{v}_{i,t}\| \|\mathbf{x}_{i,t} - \mathbf{v}_{i,t}\|, \\
      &\le \frac{\nu}{\theta} \alpha_t \beta_t  \|\hat{\mathbf{v}}_{i,t} - \mathbf{v}_{i,t}\| \sum_{j=1}^N a_{ij} \|\mathbf{x}_{i,t} - \mathbf{q}_{j,t}\|.
    \end{aligned}
  \end{equation}
  Recall that $\mathbf{v}_{i,t}, \hat{\mathbf{v}}_{i,t} \in \mathcal{X}$ and $\mathcal{X}$ is a compact set, which means that there exists some constant $C_{\mathcal{X}}$ such that $\|\hat{\mathbf{v}}_{i,t} - \mathbf{v}_{i,t}\| \le C_{\mathcal{X}}$. Similarly, $\|\mathbf{x}_{i,t} - \mathbf{q}_{j,t}\|$ can be expanded as
  \begin{equation}
    \|\mathbf{x}_{i,t} - \mathbf{q}_{j,t}\| \le \|\mathbf{x}_{i,t} - \mathbf{x}_{j,t}\| + \|\mathbf{x}_{j,t} - \mathbf{q}_{j,t}\| \le C_{\mathcal{X}} + \Delta.
  \end{equation}
  In terms of the above derivation, it follows that
  \begin{equation}
    \frac{\alpha_t}{\theta} \langle \hat{\mathbf{v}}_{i,t} - \mathbf{v}_{i,t}, \mathbf{g}_{i,t} - \mathbf{u}_{i,t} \rangle \le \frac{\nu}{\theta} C_{\mathcal{X}}(C_{\mathcal{X}} + \Delta) \alpha_t \beta_t,
  \end{equation}
  which is actually summable with respect to $t$. 
  
  As can be concluded from (\ref{eqn:varphi-xt1-expansion}), the primary challenge to obtain the iterative expression of $\varphi_{\theta}(\mathbf{x}_{i,t+1})$ is finding the connection between $\varphi_{\theta}(\mathbf{v}_{i,t})$ and $\varphi_{\theta}(\mathbf{x}_{i,t})$. Before discussing $\varphi_{\theta}(\mathbf{v}_{i,t})$ and $\varphi_{\theta}(\mathbf{x}_{i,t})$, we first put focus on $f(\hat{\mathbf{v}}_{i,t}) - f(\mathbf{v}_{i,t})$ and $\| \hat{\mathbf{v}}_{i,t} - \mathbf{v}_{i,t} \|^2$. To establish a relationship among $\mathbf{v}_{i,t}$, $\mathbf{x}_{i,t}$ and $\hat{\mathbf{v}}_{i,t}$, we rely on an auxiliary variable $\mathbf{s}_t:= \text{prox}_{\theta f} (\bar{\mathbf{x}}_t)$ which is a proximal point of $\varphi(\bar{\mathbf{x}}_t)$. 
  \begin{equation}
    \label{eqn:v-diff}
    \begin{aligned}
      & \ \frac{\rho}{2} \| \hat{\mathbf{v}}_{i,t} - \mathbf{v}_{i,t} \|^2 \\
      &= \frac{\rho}{2} \| \mathbf{v}_{i,t} - \bar{\mathbf{x}}_t + \bar{\mathbf{x}}_t - \mathbf{s}_t + \mathbf{s}_t - \hat{\mathbf{v}}_{i,t} \|^2, \\
      &\le \rho \|\mathbf{s}_{t} - \bar{\mathbf{x}}_t\|^2 + \rho \|\mathbf{v}_{i,t} - \bar{\mathbf{x}}_t + \mathbf{s}_t - \hat{\mathbf{v}}_{i,t} \|^2, \\
      &\le \rho \|\mathbf{s}_{t} - \bar{\mathbf{x}}_t\|^2 + 2\rho \left[1 + \frac{1}{(1-\theta\rho)^2} \right] \|\mathbf{v}_{i,t} - \bar{\mathbf{x}}_t\|^2.
    \end{aligned}
  \end{equation}
  $\|\mathbf{v}_{i,t} - \bar{\mathbf{x}}_t\|^2$ can be expanded as
  \begin{equation}
    \begin{aligned}
      & \ \mathbb{E} [\|\mathbf{v}_{i,t} - \bar{\mathbf{x}}_t\|^2|\mathscr{F}_t] \\
      % &= \mathbb{E} \left[ \left\| (1-\beta_t) (\mathbf{x}_{i,t} - \bar{\mathbf{x}}_t) + \beta_t \sum_{j=1}^N a_{ij} (\mathbf{q}_{j,t} - \bar{\mathbf{x}}_t)  \right\|^2 | \mathscr{F}_t \right], \\
      &\le (1-\beta_t) \|\mathbf{x}_{i,t} - \bar{\mathbf{x}}_t\|^2 + \beta_t \sum_{j=1}^N a_{ij} \mathbb{E} [ \|\mathbf{q}_{j,t} - \bar{\mathbf{x}}_t\|^2 | \mathscr{F}_t],\\
      &\le (1-\beta_t) \|\mathbf{x}_{i,t} - \bar{\mathbf{x}}_t\|^2 + \beta_t \sum_{j=1}^N a_{ij} \|\mathbf{x}_{j,t} - \bar{\mathbf{x}}_t\|^2 \\
      &\quad + \beta_t \sum_{j=1}^N a_{ij} \mathbb{E} [  \|\mathbf{x}_{j,t} - \mathbf{q}_{j,t}\|^2 | \mathscr{F}_t],
    \end{aligned}
  \end{equation}
  where by (\ref{eqn:quantization}), the last term satisfies
  \begin{equation}
    \beta_t \sum_{j=1}^N a_{ij} \mathbb{E} [\|\mathbf{x}_{j,t} - \mathbf{q}_{j,t}\|^2| \mathscr{F}_t] \le \frac{\beta_t \Delta^2}{4}.
  \end{equation}
  Similarly, $f_i(\hat{\mathbf{v}}_{i,t}) - f_i(\mathbf{v}_{i,t})$ can be written as 
  \begin{equation}
    \label{eqn:f-v-diff}
    \begin{aligned}
      & \ f_i(\hat{\mathbf{v}}_{i,t}) - f_i(\mathbf{v}_{i,t}) \\
      &= f_i(\hat{\mathbf{v}}_{i,t}) - f_i(\mathbf{s}_t) + f_i(\mathbf{s}_t) -f_i(\bar{\mathbf{x}}_t) + f_i(\bar{\mathbf{x}}_t) - f_i(\mathbf{v}_{i,t}), \\
      &\le L \| \hat{\mathbf{v}}_{i,t} - \mathbf{s}_t \| + L \| \mathbf{v}_{i,t} - \bar{\mathbf{x}}_t \| + f_i(\mathbf{s}_t) -f_i(\bar{\mathbf{x}}_t), \\
      &\le L \left( 1+\frac{1}{1-\theta \rho} \right) \| \mathbf{v}_{i,t} - \bar{\mathbf{x}}_t \| + f_i(\mathbf{s}_t) -f_i(\bar{\mathbf{x}}_t), \\
      &\le L \left( 1+\frac{1}{1-\theta \rho} \right) (1-\beta_t) \|\mathbf{x}_{i,t} - \bar{\mathbf{x}}_t\| + f_i(\mathbf{s}_t) -f_i(\bar{\mathbf{x}}_t) \\
      &+ \beta_t L \left( 1+\frac{1}{1-\theta \rho} \right) \sum_{j=1}^N a_{ij} ( \|\mathbf{x}_{j,t} - \bar{\mathbf{x}}_t\| + \| \mathbf{x}_{j,t} - \mathbf{q}_{j,t} \| ).
    \end{aligned}
  \end{equation}
  Moreover, by (\ref{eqn:quantization}), we have
  \begin{equation}
    \mathbb{E}\|\mathbf{x}_{j,t} - \mathbf{q}_{j,t}\| \le \Delta /2.
  \end{equation}
  Next we proceed to expand $\varphi_{\theta}(\mathbf{v}_{i,t})$ as
  \begin{equation*}
    \begin{aligned}
      \varphi_{\theta}(\mathbf{v}_{i,t}) &= f(\hat{\mathbf{v}}_{i,t}) + \frac{1}{2\theta} \left\| \hat{\mathbf{v}}_{i,t} - \mathbf{v}_{i,t} \right\|^2, \\
      &\le f\left( (1-\beta_t) \hat{\mathbf{x}}_{i,t} + \beta_t \sum_{j=1}^N a_{ij}\hat{\mathbf{x}}_{j,t} \right) \\
      &+ \frac{1}{2\theta} \left\| (1-\beta_t) \hat{\mathbf{x}}_{i,t} + \beta_t \sum_{j=1}^N a_{ij}\hat{\mathbf{x}}_{j,t} - \mathbf{v}_{i,t} \right\|^2,
    \end{aligned}
  \end{equation*} 
  where ``$\le$'' stems from the proximality of $\hat{\mathbf{v}}_{i,t}$. According to the definition of $\mathbf{v}_{i,t}$, we have
  \begin{equation*}
    \begin{aligned}
      & \left\| (1-\beta_t) \hat{\mathbf{x}}_{i,t} + \beta_t \sum_{j=1}^N a_{ij}\hat{\mathbf{x}}_{j,t} - \mathbf{v}_{i,t} \right\|^2 \\
      &= \left\| (1-\beta_t) (\hat{\mathbf{x}}_{i,t} - \mathbf{x}_{i,t}) + \beta_t \sum_{j=1}^N a_{ij} (\hat{\mathbf{x}}_{j,t} - \mathbf{q}_{j,t}) \right\|^2 , \\
      &\le (1-\beta_t) \|\hat{\mathbf{x}}_{i,t} - \mathbf{x}_{i,t}\|^2 + \beta_t \sum_{j=1}^N a_{ij} \|\hat{\mathbf{x}}_{j,t} - \mathbf{q}_{j,t}\|^2, \\
      &= (1-\beta_t) \|\hat{\mathbf{x}}_{i,t} - \mathbf{x}_{i,t}\|^2 + \beta_t \sum_{j=1}^N a_{ij} \|\hat{\mathbf{x}}_{j,t} - \mathbf{x}_{j,t}\|^2 \\ 
      &+ \beta_t \sum_{j=1}^N a_{ij} ( \|\mathbf{x}_{j,t} - \mathbf{q}_{j,t}\|^2 + 2\langle \hat{\mathbf{x}}_{j,t} - \mathbf{x}_{j,t}, \mathbf{x}_{j,t} - \mathbf{q}_{j,t} \rangle),
    \end{aligned}
  \end{equation*}
  where the last two terms are induced by quantization. By Lemma \ref{lma:weakly-convex-property}, $f(\cdot)$ can be written as
  \begin{equation}
    \begin{aligned}
      & f\left( (1-\beta_t) \hat{\mathbf{x}}_{i,t} + \beta_t \sum_{j=1}^N a_{ij}\hat{\mathbf{x}}_{j,t} \right)\\
      &\le (1-\beta_t) f(\hat{\mathbf{x}}_{i,t}) + \beta_t \sum_{j=1}^N a_{ij} f(\hat{\mathbf{x}}_{j,t}) \\
      &+ \frac{\rho}{2} \sum_{j=0}^{N-1} \sum_{l=j+1}^N c_{ij} c_{il} \|\hat{\mathbf{x}}_{j,t} - \hat{\mathbf{x}}_{l,t}\|^2.
    \end{aligned}
  \end{equation}
  % Moreover, it follows that
  % \begin{equation}
  %   \begin{aligned}
  %     \ f\left( (1-\beta_t) \hat{\mathbf{x}}_{i,t} + \beta_t \sum_{j=1}^N a_{ij}\hat{\mathbf{x}}_{j,t} \right) \le \frac{1}{N} \sum_{i=1}^N f(\hat{\mathbf{x}}_{i,t}) \\
  %     + \frac{\rho}{2(1-\theta \rho)} \sum_{i=1}^N \sum_{j=0}^{N-1} \sum_{l=j+1}^N c_{ij} c_{il} \|\hat{\mathbf{x}}_{j,t} - \hat{\mathbf{x}}_{l,t}\|^2.
  %   \end{aligned}
  % \end{equation}
  Summing the terms above, we obtain
  \begin{equation}
    \label{eqn:varphi-v-expansion}
    \begin{aligned}
      & \varphi_{\theta}(\mathbf{v}_{i,t}) \\
      &\le (1-\beta_t) \varphi_{\theta}(\mathbf{x}_{i,t}) + \beta_t \sum_{j=1}^N a_{ij} \varphi_{\theta}(\mathbf{x}_{j,t}) \\
      &+ \beta_t \sum_{j=1}^N a_{ij} ( \|\mathbf{x}_{j,t} - \mathbf{q}_{j,t}\|^2 + 2\langle \hat{\mathbf{x}}_{j,t} - \mathbf{x}_{j,t}, \mathbf{x}_{j,t} - \mathbf{q}_{j,t} \rangle) \\
      &+ \frac{\rho}{2} \sum_{j=0}^{N-1} \sum_{l=j+1}^N c_{ij} c_{il} \|\hat{\mathbf{x}}_{j,t} - \hat{\mathbf{x}}_{l,t}\|^2,
    \end{aligned}
  \end{equation}
  where $c_{ij}$ is given by
  \begin{equation}
    \label{eqn:cij}
    c_{ij} = \left\{
    \begin{aligned}
      1-\beta_t, && j=0, \\
      \beta_t a_{ij}, && j>0,\\
    \end{aligned}
    \right.
  \end{equation}
  and $\hat{\mathbf{x}}_{0,t}$ is defined as $\hat{\mathbf{x}}_{i,t}$. Note that (\ref{eqn:varphi-v-expansion}) bridges the gap between $\varphi_{\theta}(\mathbf{v}_{i,t})$ and $\varphi_{\theta}(\mathbf{x}_{j,t}), \forall j$. Taking expectation at both sides to evaluate the quantization error, we obtain
  \begin{equation}
    \label{eqn:varphi-v-x}
    \begin{aligned}
      & \mathbb{E}[\varphi_{\theta}(\mathbf{v}_{i,t})] \\
      &\le (1-\beta_t) \mathbb{E} [\varphi_{\theta}(\mathbf{x}_{i,t})] + \beta_t \sum_{j=1}^N a_{ij} \varphi_{\theta}(\mathbf{x}_{j,t}) + \frac{\beta_t \Delta^2}{4} \\
      &\quad + \frac{\rho}{2} \sum_{j=0}^{N-1} \sum_{l=j+1}^N c_{ij} c_{il} \|\hat{\mathbf{x}}_{j,t} - \hat{\mathbf{x}}_{l,t}\|^2.
    \end{aligned}
  \end{equation}
  Observe that the function $\mathbf{x}\mapsto f_i(\mathbf{x})+\frac{1}{2\theta} \|\mathbf{x} - \bar{\mathbf{x}}_t\|^2$ is strongly convex with parameter $\frac{1}{\theta}-\rho$ if $\theta < \frac{2}{3\rho}$. It follows that
  \begin{equation}
    \label{eqn:f-difference-st-xt}
    \begin{aligned}
      & \ f_i(\mathbf{s}_t) - f_i(\bar{\mathbf{x}}_t) + \rho \|\mathbf{s}_t-\bar{\mathbf{x}}_t\|^2 \\
      &\le f_i(\mathbf{s}_t) + \frac{1}{2\theta} \|\mathbf{s}_t - \bar{\mathbf{x}}_t\|^2 - \left[f_i(\bar{\mathbf{x}}_t) + \frac{1}{2\theta} \|\bar{\mathbf{x}}_t - \bar{\mathbf{x}}_t \|^2 \right] \\
      &\quad + \left(\rho - \frac{1}{2\theta}\right) \|\mathbf{s}_t - \bar{\mathbf{x}}_t\|^2, \\
      &\le \left( \frac{3}{2}\rho - \frac{1}{\theta} \right) \|\mathbf{s}_t - \bar{\mathbf{x}}_t\|^2 \le 0.
    \end{aligned}
  \end{equation}
  Substituting (\ref{eqn:v-diff}), (\ref{eqn:f-v-diff}), (\ref{eqn:varphi-v-x}) and (\ref{eqn:f-difference-st-xt}) into (\ref{eqn:varphi-xt1-expansion}) and letting $\bar{\varphi}^t_\theta :=\frac{1}{N} \sum_{i=1}^N \varphi_\theta (\mathbf{x}_{i,t})$, we obtain
  \begin{equation}
    \label{eqn:varphi-bar-iter1}
    \begin{aligned}
      & \ \mathbb{E}[\bar{\varphi}^{t+1}_\theta | \mathscr{F}_t] \\
      &\le (1-\beta_t) \mathbb{E}[\bar{\varphi}^t_\theta] + \frac{1}{N}\beta_t \sum_{i=1}^N \sum_{j=1}^N a_{ij} \mathbb{E}[\varphi_\theta (\mathbf{x}_{j,t})] \\
      &+ \frac{\rho}{2N} \sum_{i=1}^N \sum_{j=0}^{N-1} \sum_{l=j+1}^N c_{ij} c_{il} \|\hat{\mathbf{x}}_{j,t} - \hat{\mathbf{x}}_{l,t}\|^2 \\
      &+ \frac{L\alpha_t}{N\theta}\left(1+\frac{1}{1-\theta \rho} \right) \sum_{i=1}^N \|\mathbf{x}_{i,t} - \bar{\mathbf{x}}_t\| \\
      &+ \frac{2\rho\alpha_t}{N\theta} \left[1+\frac{1}{(1-\theta\rho)^2}\right]\sum_{i=1}^N \|\mathbf{x}_{i,t} - \bar{\mathbf{x}}_t\|^2 \\
      &+ \frac{\beta_t \alpha_t \rho \Delta^2}{2N\theta} \left[1+\frac{1}{(1-\theta\rho)^2}\right] + \frac{\alpha_t^2 L^2}{2\theta} \\
      &+ \frac{\alpha_t \beta_t L \Delta}{2N\theta} \left(1+\frac{1}{1-\theta \rho}\right) + \frac{\nu}{\theta} C_{\mathcal{X}}(C_{\mathcal{X}} + \Delta) \alpha_t \beta_t.
    \end{aligned}
  \end{equation}
  Moreover, we have
  \begin{equation}
    \label{eqn:varphi-martingale}
    \mathbb{E}[\bar{\varphi}^{t+1}_\theta| \mathscr{F}_t] \le \bar{\varphi}^{t}_\theta + b_t + q_t + \frac{\alpha_t^2 L^2}{2\theta},
  \end{equation}
  where $b_t$ consists of the terms containing $\|\mathbf{x}_{i,t} - \bar{\mathbf{x}}_t\|$, and $q_t$ represents the terms containing quantization error. It is straightforward to verify that
  \begin{equation}
    \sum_{t=1}^\infty \frac{\alpha_t^2}{\beta_t}<\infty \Longrightarrow \sum_{t=1}^\infty q_t < \infty.
  \end{equation}
  Next, we continue to derive $b_t$. According to the definition of $c_{ij}$ as (\ref{eqn:cij}), we have
  \begin{equation}
    c_{ij} c_{il} \|\hat{\mathbf{x}}_{j,t} - \hat{\mathbf{x}}_{l,t}\|^2 \le \frac{2a_{il}\beta_t}{(1-\theta \rho)^2} ( \|\mathbf{x}_{j,t} - \bar{\mathbf{x}}_t\|^2 + \|\mathbf{x}_{l,t} - \bar{\mathbf{x}}_t\|^2),
  \end{equation}
  which by Lemma \ref{lma:x-deviation} yields $\sum_{t=1}^\infty c_{ij} c_{il} \|\hat{\mathbf{x}}_{j,t} - \hat{\mathbf{x}}_{l,t}\|^2 < \infty$. In addition, since $\sum_{t=1}^\infty \alpha_t^2/\beta_t < \infty$, $\sum_{t=1}^\infty \alpha_t = \infty$, and $\sum_{t=1}^\infty \beta_t \|\mathbf{x}_{i,t} - \bar{\mathbf{x}}_t\|^2 < \infty$, we conclude
  \begin{equation}
    \alpha_t \|\mathbf{x}_{i,t} - \bar{\mathbf{x}}_t\|^2 < \infty.
  \end{equation} 
  Moreover, $\alpha_t \|\mathbf{x}_{i,t} - \bar{\mathbf{x}}_t\|$ can be written as
  \begin{equation*}
    \begin{aligned}
      \sum_{t=1}^\infty \alpha_t \|\mathbf{x}_{i,t} - \bar{\mathbf{x}}_t\| &= \sum_{t=1}^\infty  \frac{\alpha_t}{\sqrt{\beta_t}} \sqrt{\beta_t} \|\mathbf{x}_{i,t} - \bar{\mathbf{x}}_t\|, \\
      &\le  \frac{1}{2} \sum_{t=1}^\infty  \left( \frac{\alpha_t^2}{\beta_t} + \beta_t \|\mathbf{x}_{i,t} - \bar{\mathbf{x}}_t\|^2 \right) < \infty.
    \end{aligned}
  \end{equation*}
  Therefore, we have $\sum_{t=1}^\infty b_t < \infty$. 

  Since $f$ is assumed to be lower bounded, $\bar{\varphi}^{t}_\theta$ is also lower bounded, i.e., there exists $\varphi_\theta^\dagger = \inf \bar{\varphi}_\theta^t$ such that
  \begin{equation}
    \mathbb{E}[\bar{\varphi}^{t+1}_\theta| \mathscr{F}_t] - \varphi_\theta^\dagger \le \bar{\varphi}^{t}_\theta - \varphi_\theta^\dagger + b_t + q_t + \frac{\alpha_t^2 L^2}{2\theta},
  \end{equation}
  which, by the convergence theorem of almost supermartingales \cite{1971SuperMartingale} and in consideration of the finiteness of $\sum_{t=1}^\infty b_t$, $\sum_{t=1}^\infty q_t$ and $\sum_{t=1}^\infty \alpha_t^2$, indicates that there exists $\bar{\varphi}^*_\theta$ satisfying
  \begin{equation}
    \label{eqn:varphi-bar-convergence}
    \lim_{t\rightarrow \infty} \bar{\varphi}^{t}_\theta = \bar{\varphi}^*_\theta, \quad \text{a.s.}
  \end{equation}
  Furthermore, since $\varphi_\theta(\cdot)$ is continuously differentiable \cite[Lemma 2.2]{2019WeaklyConvex} and $\lim_{t\rightarrow \infty}\|\mathbf{x}_{i,t} - \bar{\mathbf{x}}_t\| = 0$, it follows that
  \begin{equation}
    \label{eqn:varphi-average-convergence}
    \mathbb{E} | \varphi_\theta(\mathbf{x}_{i,t}) - \varphi_\theta(\bar{\mathbf{x}}_t) |^2 \rightarrow 0.
  \end{equation}
  Therefore, the limitation of $\varphi_\theta (\bar{\mathbf{x}}_t)$ can be established as
  \begin{equation}
    \label{eqn:varphi-xbar-convergence}
    \begin{aligned}
      \mathbb{E}|\varphi_\theta (\bar{\mathbf{x}}_t) - \bar{\varphi}_\theta^t|^2 &= \mathbb{E} \left| \frac{1}{N} \sum_{i=1}^N \varphi_\theta (\mathbf{x}_{i,t}) - \varphi_\theta (\bar{\mathbf{x}}_t) \right|^2, \\
      &\le \frac{1}{N} \sum_{i=1}^N \mathbb{E} | \varphi_\theta (\mathbf{x}_{i,t}) - \varphi_\theta (\bar{\mathbf{x}}_t) |^2,
    \end{aligned}
  \end{equation}
  which means that $\varphi_\theta (\bar{\mathbf{x}}_t) \rightarrow \bar{\varphi}^*_\theta$, a.s. Combining (\ref{eqn:varphi-bar-convergence}), (\ref{eqn:varphi-average-convergence}) and (\ref{eqn:varphi-xbar-convergence}), we conclude that $\varphi_{\theta}(\mathbf{x}_{i,t}) \rightarrow \bar{\varphi}_\theta^*$, $\forall i\in \mathcal{V}$.

  Next we proceed to show that the limit $\bar{\varphi}_\theta^*$ is actually the extreme point of $\varphi_\theta(\bar{\mathbf{x}}_t)$. Recall that $\varphi_\theta(\cdot)$ is continously differentiable. Therefore, it is sufficient to show the gradient of $\varphi_\theta(\bar{\mathbf{x}}_t)$ diminishes with respect to iteration $t$, i.e., $\|\nabla \varphi_\theta(\bar{\mathbf{x}}_t)\| \rightarrow 0$. $\nabla \varphi_\theta(\bar{\mathbf{x}}_t)$ can be written as \cite[Lemma 2.2]{2019WeaklyConvex}
  \begin{equation}
    \nabla \varphi_\theta(\bar{\mathbf{x}}_t) = \frac{1}{\theta} (\bar{\mathbf{x}}_t - \mathbf{s}_t).
  \end{equation}
  Rearranging (\ref{eqn:f-difference-st-xt}), (\ref{eqn:varphi-bar-iter1}) and (\ref{eqn:varphi-martingale}), we obtain
  \begin{equation*}
    \frac{\alpha_t}{\theta} \left(\frac{1}{\theta} - \frac{3}{2}\rho\right) \|\mathbf{s}_t - \mathbf{x}_t\|^2 \le \bar{\varphi}_\theta^t - \bar{\varphi}_\theta^{t+1} + b_t + q_t + \frac{\alpha_t^2L^2}{2\theta}.  
  \end{equation*}
  Unfolding the recursion yields for some $T>0$
  \begin{equation}
    \label{eqn:recursion-varphi-T}
    \bar{\varphi}_\theta^T \le \bar{\varphi}_\theta^0 + \sum_{t=0}^T \varsigma_t - \left(1-\frac{3}{2}\theta \rho\right) \sum_{t=0}^T \alpha_t \|\nabla \varphi_\theta(\bar{\mathbf{x}}_t)\|^2,
  \end{equation}
  where $\varsigma_t = \frac{\alpha_t^2 L^2}{2\theta} + b_t + q_t$ denotes the summable term. Lower bounding the left hand side by $\varphi_\theta^\dagger$, we have
  \begin{equation*}
    \left(1-\frac{3}{2}\theta \rho\right) \sum_{t=0}^T \alpha_t \mathbb{E} \|\nabla \varphi_\theta(\bar{\mathbf{x}}_t)\|^2  \le \sum_{t=0}^T \mathbb{E}[\varsigma_t] + \bar{\varphi}_\theta^0 - \varphi_\theta^\dagger.
  \end{equation*}
  Rearranging both sides, we obtain
  \begin{equation*}
    \inf_{0\le t\le T} \mathbb{E} \|\nabla \varphi_\theta (\bar{\mathbf{x}}_t)\|^2 \le \frac{2}{2-3\theta \rho} \frac{\sum_{t=0}^T \mathbb{E}[\varsigma_t] + \bar{\varphi}_\theta^0 - \varphi_\theta^\dagger}{\sum_{t=0}^T \alpha_t}.
  \end{equation*}
  Since $\sum_{t=0}^\infty \mathbb{E}[\varsigma_t] < \infty$, and $\sum_{t=0}^\infty \alpha_t = \infty$, it follows that
  \begin{equation}
    \liminf_{t\rightarrow \infty} \mathbb{E} \|\nabla \varphi_\theta (\bar{\mathbf{x}}_t)\|^2 = 0.
  \end{equation}
  Therefore, there exists a subsequence $\{\bar{\mathbf{x}}_{k_\ell}\}$ such that
  \begin{equation}
    \lim_{\ell \rightarrow \infty} \mathbb{E} \|\nabla \varphi_\theta (\bar{\mathbf{x}}_{k_\ell})\|^2 = 0,
  \end{equation}
  which incdicates that
  \begin{equation}
    \varphi_\theta (\bar{\mathbf{x}}_{k_\ell}) \rightarrow \tilde{\varphi}_\theta = \min_{\mathbf{y}} \varphi_\theta(\mathbf{y}).
  \end{equation}
  Since $\varphi_\theta$ is smooth and convex, we conclude $\bar{\varphi}^*_\theta=\tilde{\varphi}_\theta$ and $\|\nabla \varphi_\theta (\bar{\mathbf{x}}_t) \| \rightarrow 0$ by the convergence of $\{\varphi_\theta (\bar{\mathbf{x}}_t)\}$. 
\end{proof}

\bibliographystyle{IEEEtran}
\bibliography{IEEEabrv,ref}

\end{document}